\documentclass[acmsmall, screen]{acmart}

\AtBeginDocument{%
  }

\setcopyright{acmlicensed}

\usepackage{longtable}

\usepackage{amssymb} 
\usepackage{array} 

\usepackage{xspace}
\usepackage{setspace}
\usepackage{tcolorbox}
\usepackage{booktabs} 
\usepackage{ragged2e}
\usepackage{multirow}
\usepackage{graphicx}
\usepackage{adjustbox}
\usepackage{tabularx}
\newcolumntype{C}[1]{>{\centering\arraybackslash}p{#1}}
\usepackage{colortbl}
\usepackage{graphicx}  
\usepackage{float}  
\usepackage{subfigure}  
\usepackage{color}
\usepackage{arydshln} 
\usepackage{caption}
\usepackage{wasysym}
\usepackage{tikz}
\usepackage{makecell}
\usepackage{enumitem}

\usepackage{microtype}
\newcommand*\circled[1]{\tikz[baseline=(char.base)]{
            \node[shape=circle,draw,inner sep=0.5pt] (char) {#1};}}

\newcolumntype{M}[1]{>{\centering\arraybackslash}m{#1}}

\newcommand{\ie}[0]{\textit{i.e.,}\xspace}
\newcommand{\eg}[0]{\textit{e.g.,}\xspace}
\newcommand{\etc}[0]{\textit{etc.}\xspace}

\newcommand{\todo}[1]{\textcolor{black}{#1}}

\begin{document}

\title{A Survey of Fuzzing Open-Source Operating Systems}

\author{Kun Hu}
\email{huk23@m.fudan.edu.cn}
\affiliation{%
  \department{College of Computer Science and Artificial Intelligence}
  \institution{Fudan University}
  \state{Shanghai}
  \country{China}
  \postcode{200433}
}

\author{Qicai Chen}
\email{qcchen23@m.fudan.edu.cn}
\affiliation{%
  \department{College of Computer Science and Artificial Intelligence}
  \institution{Fudan University}
  \state{Shanghai}
  \country{China}
  \postcode{200433}
}

\author{Wenzhuo Zhang}
\email{wzzhang24@m.fudan.edu.cn}
\affiliation{%
  \department{College of Computer Science and Artificial Intelligence}
  \institution{Fudan University}
  \state{Shanghai}
  \country{China}
  \postcode{200433}
}

\author{Zilong Lu}
\email{zllu23@m.fudan.edu.cn}
\affiliation{%
  \department{College of Computer Science and Artificial Intelligence}
  \institution{Fudan University}
  \state{Shanghai}
  \country{China}
  \postcode{200433}
}

\author{Bihuan Chen}
\email{bhchen@fudan.edu.cn}
\authornote{Bihuan Chen is the corresponding author.}
\affiliation{%
  \department{College of Computer Science and Artificial Intelligence}
  \institution{Fudan University}
  \state{Shanghai}
  \country{China}
  \postcode{200433}
}


\author{You Lu}
\email{ylu24@m.fudan.edu.cn}
\affiliation{%
  \department{College of Computer Science and Artificial Intelligence}
  \institution{Fudan University}
  \state{Shanghai}
  \country{China}
  \postcode{200433}
}

\author{Haowen Jiang}
\email{hwjiang23@m.fudan.edu.cn}
\affiliation{%
  \department{College of Computer Science and Artificial Intelligence}
  \institution{Fudan University}
  \state{Shanghai}
  \country{China}
  \postcode{200433}
}

\author{Bingkun Sun}
\email{bksun21@m.fudan.edu.cn}
\affiliation{%
  \department{College of Computer Science and Artificial Intelligence}
  \institution{Fudan University}
  \state{Shanghai}
  \country{China}
  \postcode{200433}
}

\author{Xin Peng}
\email{pengxin@fudan.edu.cn}
\affiliation{%
  \department{College of Computer Science and Artificial Intelligence}
  \institution{Fudan University}
  \state{Shanghai}
  \country{China}
  \postcode{200433}
}

\author{Wenyun Zhao}
\email{wyzhao@fudan.edu.cn}
\affiliation{%
  \department{College of Computer Science and Artificial Intelligence}
  \institution{Fudan University}
  \state{Shanghai}
  \country{China}
  \postcode{200433}
}



\renewcommand{\shortauthors}{Hu et al.}


\begin{abstract}

Vulnerabilities in open-source operating systems (OSs)  pose substantial security risks to software systems, making their detection crucial. While fuzzing has been an effective vulnerability detection technique in various domains, OS fuzzing (OSF) faces unique challenges due to OS complexity and multi-layered interaction, and has not been comprehensively reviewed. Therefore, this work systematically surveys the state-of-the-art OSF techniques, categorizes them based on the general fuzzing process, and investigates challenges specific to kernel, file system, driver, and hypervisor fuzzing. Finally, future research directions for OSF are discussed. 

\end{abstract}

%




\maketitle


\section{Introduction}

Operating systems (OSs) play a crucial role in the operation of modern computer systems.~They~are responsible for managing the hardware and software resources of a computer. 
They serve~as~the cornerstone for maximizing hardware resource utilization and ensuring software system stability. Specifically, open-source OSs, such as Linux, Android, FreeRTOS, Zephyr, \etc, have become~particularly favored in most fields like autonomous driving, 
robotics, 
cloud services, 
and Web of Things 
for the advantages of transparency, customizability, and community-driven innovation, leading to their anticipated dominance in future computer systems.

However, as the scale of open-source OS codebases continues to expand, the security threats they pose have become increasingly alarming, raising significant concerns about the secure operation of software systems. According to National Vulnerability Database (NVD) \cite{NVD},~more~than~9,300~vulnerabilities have been discovered in the kernels of mainstream open-source OSs~(including Linux, Android, FreeBSD, OpenBSD and Zephyr) since 2004. In particular, the number of vulnerabilities surges to 3,300 in the single year of 2024, an increase that is 10 times higher than the previous~year. 
Similarly, the Common Vulnerabilities and Exposures (CVE) \cite{cve} has documented at least 7,600 vulnerabilities in the Linux kernel to date. Among them, NVD has disclosed at least 1,566 high-severity vulnerabilities and 156 critical vulnerabilities that require immediate remediation. 


OS-level vulnerabilities are particularly concerning when compared to those in user-level applications. This is because such vulnerabilities can be further exploited, potentially allowing attackers to gain complete control over the OSs \cite{Xu2015FromCT, Zhang2015AndroidRA,chen2019slake}, leading to irreversible consequences. The staggering number of these vulnerabilities and the malicious outcomes they have caused have~attracted significant attention from security researchers. As a result, there is a strong and growing~interest in developing effective and efficient techniques to identify and mitigate these potential vulnerabilities, thereby aiding the continuous evolution of open-source OSs. One of the widely used techniques is fuzzing, which was first introduced by Miller et al. \cite{Miller1990AnES} in 1990, and has achieved notable~success across various domains, \eg compilers \cite{Yang2023WhiteboxCF}, interpreters \cite{Holler2012FuzzingWC}, and open-source software~\cite{Serebryany2017OSSFuzzG}.

Fuzzing, also known as fuzz testing, is a technique that involves feeding semi-randomly generated test cases as inputs to the program under test (PUT) to trigger program paths that may contain software vulnerabilities. Over the past decade, fuzzing has become capable of effectively testing complex OS code. This progress has received widespread attention from researchers, who aim to enhance the depth and breadth of OS fuzzing ({OSF}) by incorporating cutting-edge techniques~such~as program analysis and deep learning. However, compared to traditional fuzzing, the complex domain knowledge involved in OS makes developing effective and efficient OSF particularly~challenging. It not only requires focusing on advancements in fuzzing techniques, but also demands consideration of the inherent complexity and multi-layered interaction of OS. Typically, survey papers play~a~key role in advancing this field by providing a comprehensive review of the OSF techniques as well as summarizing existing challenges and pinpointing potential directions.

However, to date, there has been no systematic review of OSF. While there are some survey~papers on traditional software fuzzing \cite{Eisele2022EmbeddedFA,Yun2022FuzzingOE,Zhu2022FuzzingAS,Mans2018TheAS,Li2018FuzzingAS,Liang2018FuzzingSO,Zhang2018SurveyOD,Mallissery2023DemystifyTF,bohme2020fuzzing,Godefroid2020}, they do not systematically introduce the general steps of OSF, and they also do not highlight the unique challenges that~OSF faces compared to traditional software fuzzing. Therefore, to bridge the gap, a systematic review of the state-of-the-art OSF is essential, aiming to provide a comprehensive guideline on this topic.


For the scope of this survey, after conducting a systematic search of state-of-the-art OSF, we observed that the PUT is typically classified into \textit{kernel}, \textit{file system}, \textit{driver}, and \textit{hypervisor} because each of these OS layers present distinct challenges in fuzzing. Specifically, in a highly heterogeneous hardware environment (\eg for intelligent vehicles), achieving unified resource management within the OS requires the support from \textit{hypervisor}; and vulnerabilities from \textit{hypervisor} can be exploited~to launch malicious attacks against the host OS. Therefore, to ensure the completeness of our survey, we also included \textit{hypervisor}, as a component of a generalized OS, within our scope. 


To systematically review OSF, we designed a comprehensive search process to identify existing high-quality research materials (see Section \ref{Section2}), including \todo{60} state-of-the-art OSF papers and~\todo{4}~open-source OSF tools. Through a thorough analysis of these materials, we uncovered the rising trend in OSF research and explained the reasons behind it. Then, we introduced the distinctive features~of fuzzing techniques for the four OS layers (\ie kernel, file system, driver, and hypervisor) from~a high-level perspective, and summarized a general workflow of OSF, consisting of three core modules, \ie \textit{input}, \textit{fuzzing engine}, and \textit{running~environment}~(see~Section \ref{Section3}). 
Next, we conducted a comprehensive review of the state-of-the-art OSF with respect to seven key steps in the three~core~modules~(see Section \ref{Section4}), aiming~to~systematically sort out the technological advancements in OSF and the associated complex domain knowledge. Meanwhile, we summarized the unique problems encountered by the four OS layers during fuzzing as well as the corresponding solutions (see Section \ref{Section5}). 
Finally, based on our findings, we provided \todo{four} future research directions in the OSF field (see Section \ref{Section7}).

\section{Collection Strategy and Result}\label{Section2}


Following \cite{garousi2016systematic}, we use a scientific and effective collection strategy (see Section~\ref{Collection Methodology}) and present~a detailed analysis of the collection result (see Section~\ref{Results Analysis}) to systematically review OSF. 


\subsection{Collection Strategy}\label{Collection Methodology}

\begin{figure}[!t]
  \includegraphics[width=0.91\linewidth]{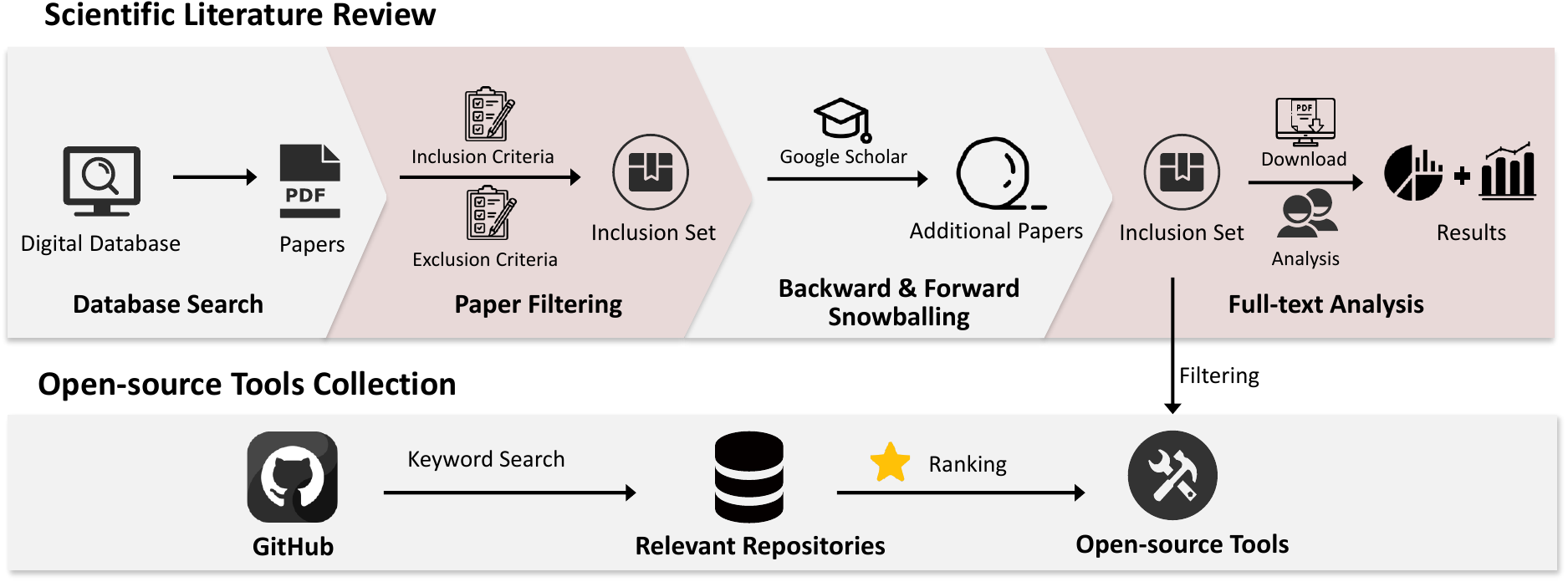}
  \vspace{-5pt}
  \caption{Our Collection Strategy}
  \label{img:collection_methodology}
\end{figure}

Figure~\ref{img:collection_methodology} shows our strategy to collect relevant works, assess their quality, and keep updated with~the latest publications. We both review scientific literature and collect open-source tools for OSF. 

\subsubsection{Scientific Literature Review}



We set the temporal scope of our survey to cover the period~from the earliest relevant papers or tools in this field to the present, ranging from January 2015~to~August 2024 (\ie around 10 years). The detailed steps are as follows.


\textbf{Database Search.} 
This step aims to find the potential relevant papers by searching electronic databases. Specifically, we select ACM Digital Library, IEEE~Xplore, DBLP and Semantic Scholar~as our databases, which are popular bibliography databases containing a comprehensive list of research venues in computer science. 
Initially, employing ``Operating System Fuzzing'' as the sole keyword fails to fully capture the existing literature about OSF. Keywords need to focus on the application~of fuzzing to operating systems (\eg Linux, Android, FreeBSD, \etc).  In addition, keywords should cover the main tested objects in open-source operating systems, such as kernel, file system, driver, \etc We optimize the search keywords in an iterative manner for the purpose of collecting as many related papers as possible. Our final search keywords are reported as follows. Moreover, our search targets titles, abstracts, and keywords of the papers, since these parts often convey the theme of a paper. Finally, we obtain a total of \todo{56} candidate papers during \emph{database search}. 

\begin{tcolorbox}[size=fbox, opacityfill=0.15]
\small
  (``Linux'' OR ``Android'' OR ``FreeBSD'' OR ``OpenBSD'' OR ``Zephyr'' OR ``Open-source Operating System'')\\
 AND  (``Kernel'' OR ``File system'' OR ``Driver'' OR ``Hypervisor'')  AND  (``Fuzzing'' OR ``OSF'')
\end{tcolorbox}


\textbf{Paper Filtering.} 
We perform a manual assessment on the \todo{56} candidate papers obtained from~our \emph{database search} to ensure their relevance and quality. Specifically, to determine whether each~candidate paper is relevant to OSF and has high quality, we analyze the abstracts and introductions of these papers, following the inclusion and exclusion criteria formulated as follows. We use two \textit{inclusion criteria}, i.e., \textbf{IC1}: papers that introduce the process of OSF, and \textbf{IC2}: papers that propose a technique of OSF. We use four \textit{exclusion criteria}, i.e., \textbf{EC1}: survey papers or summary papers, \textbf{EC2}: papers that do not target fuzzing, \textbf{EC3}: papers that do not focus on fuzzing operating system and its components, and~\textbf{EC4}: papers that have not been published in top-tier conferences or journals. Specifically, for \textbf{EC1}, such survey papers are discussed in Section~\ref{Section3-1} for a comparison with~our survey; 
and for \textbf{EC4}, we discuss these top-tier conferences and journals in Section~\ref{Results Analysis}. Through our manual assessment, we remove \todo{13} papers, resulting in \todo{43} papers.

\textbf{Backward \& Forward Snowballing.} To reduce the risk of missing relevant papers, we perform both backward and forward snowballing \cite{wohlin2014guidelines} on the \todo{43} papers. In backward snowballing,~we~check the references in these papers to obtain candidate papers, while in forward snowballing, we use Google Scholar to locate candidate papers that cite these papers. For these candidate papers obtained by snowballing, we also apply the same inclusion and exclusion criteria to identify relevant papers. Finally, we add \todo{17} new relevant papers, resulting in a final set of \todo{60} papers.

\textbf{Full-Text Analysis.} We download all the resulting \todo{60} papers, and conduct a full-text analysis~to
 identify the fuzzing target (\ie the operating system and its components) and the proposed fuzzing technique. After reading all these papers, we classify them to form our survey (see Section~\ref{Section4} and \ref{Section5}).

\subsubsection{Open-Source Tools Collection} Some open-source tools of OSF have not been published~in~academic papers, and these tools should not be ignored. Therefore, we use the same search keywords to collect open-source tools whose stars are more than 400 stars on GitHub. We eliminate tools that have been published in academic papers, and select \todo{4} additional open-source tools.


\begin{figure*}[!t]
  \centering
  \subfigure[Distribution across Publication Venues]{
    \includegraphics[width=0.42\linewidth]{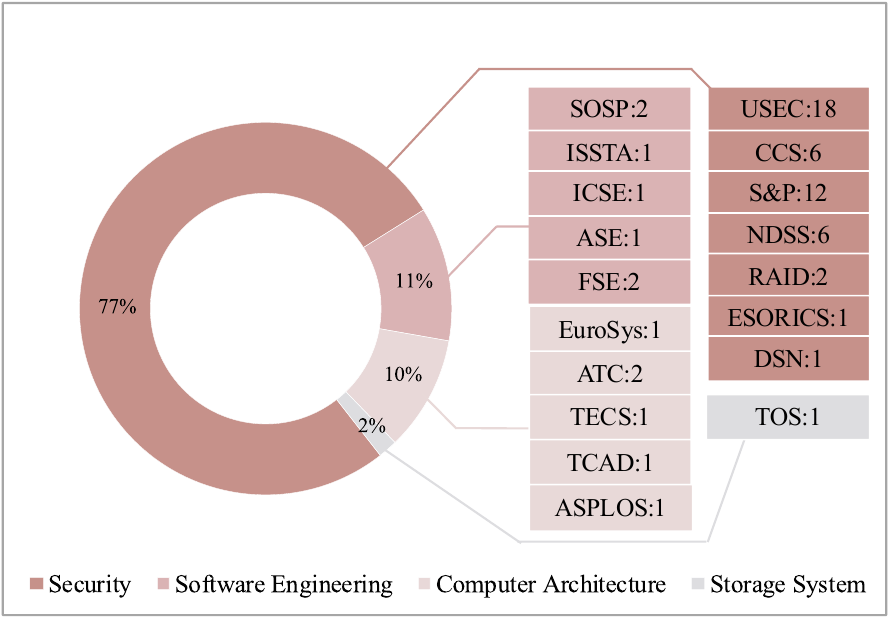}\label{fig:publication_venues}}
  \hspace{20pt}
  \subfigure[Distribution Across Publication Years and Layers]{
      \includegraphics[width=0.428\textwidth]{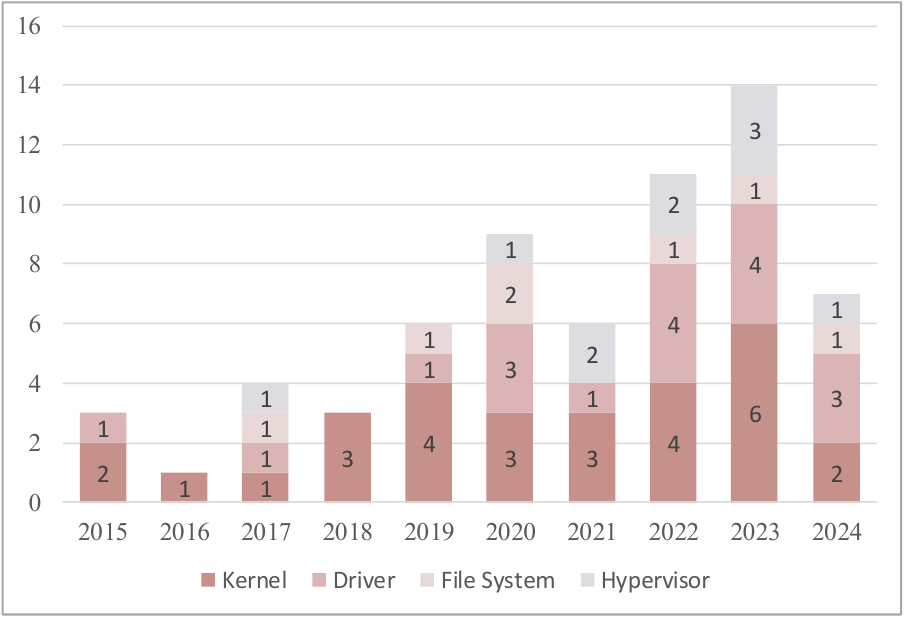}\label{fig:publication_years}}
      \vspace{-10pt}
  \caption{The Analysis Results of Paper Collection}
\end{figure*}

\subsection{Collection Result Analysis}\label{Results Analysis}

We analyze the collected papers from three perspectives, \ie the publication venues, the publication years, and the target OS layers. 

\textbf{Publication Venues.} Figure~\ref{fig:publication_venues} shows the distribution of all papers across the publication venues. The \todo{60} papers are published across \todo{18} top-tier venues in four domains, \ie security,~software engineering, computer architecture, and computer storage systems. Specifically, (\romannumeral1)~most~of~the papers, up to~\todo{77\%}, are published in security venues such as \emph{USENIX Security Symposium (USEC)}, \emph{ACM~Conference on Computer and Communications Security (CCS)}, \emph{IEEE Symposium on Security and Privacy (S\&P)}, and \emph{Network and Distributed System Security Symposium (NDSS)}; (\romannumeral2)~\todo{11\%} are published in software engineering venues such as \emph{ACM Symposium on Operating Systems Principles (SOSP)}, \emph{International Symposium on Software Testing and Analysis (ISSTA)}, and \emph{International Conference on Software Engineering (ICSE)}; (\romannumeral3)~there are \todo{5} papers, accounting for~\todo{10\%}, published in computer architecture venues, one each in \emph{European Conference on Computer Systems (EuroSys)}, \emph{USENIX Annual Technical Conference (ATC)}, \emph{ACM Transactions on Embedded Computing Systems (TECS)}, \emph{IEEE Transactions on Computer-Aided Design of Integrated Circuits and Systems (TCAD)}, and \emph{International Conference on Architectural Support for Programming Languages and Operating Systems (ASPLOS)}; and (\romannumeral4)~since operating system is related to storage system, there is~\todo{1} paper published~in computer storage system venues (\ie \emph{ACM Transactions on Storage (TOS)}). It can be conclude that application of fuzzing techniques to operating systems spans multiple fields~of~computer science.


\textbf{Publication Years and Target OS Layers.}  Figure~\ref{fig:publication_years} presents the number of papers published in each year as well as the distribution across the target OS layers in each year. Overall, the~number of OSF papers shows a general ascending trend from 2015 to 2024. It is evident that interest~in~OSF has been escalating year by year, which indicates increased attention in leveraging fuzzing~to~uncover deeper security vulnerabilities in open-source OS. Notice that since the papers from 2024 have not been fully surveyed yet, Figure~\ref{fig:publication_years} only shows the number of OSF publications till~August~2024. Moreover, most papers focus on the fuzzing of kernel. Particularly, Google’s fuzzing framework Syzkaller~\cite{Syzkaller} launched in 2015. It provides researchers with a foundational kernel fuzzing engine, and has made a substantial impact on kernel fuzzing. However, using kernel fuzzing interfaces~(\ie between user and kernel space) fails to uncover deep security vulnerabilities in other OS layers (\eg verification chain checks in driver). Therefore, an increasing number of papers realized the necessity of fuzzing other OS layers (\ie file systems, drivers, and hypervisors) since 2020.





\section{Overview of OSF}\label{Section3}

After analyzing existing fuzzing surveys, we employ PUT-based classification for a systematic~review of OSF (Section~\ref{Section3-1}). Following this classification, we introduce the main tasks of the four OS layer fuzzing (Section~\ref{Section3-2}), and provide a high-level overview of the general workflow of OSF (Section~\ref{Section3-3}).

\subsection{Classification Dimension} \label{Section3-1}

Existing fuzzing surveys classify the literature by one of the four dimensions, \ie
1) the amount~of~information the fuzzer requires or uses (black-, white-, and gray-box)~\cite{Mans2018TheAS, Li2018FuzzingAS, Liang2018FuzzingSO, Godefroid2020, bohme2020fuzzing}, 
2) the~strategy employed for seed update (mutation- and generation-based)~\cite{Saavedra2019ARO, schumilo2020hyper, schumilo2021nyx, myung2022mundofuzz, pan2021V-shuttle, cesarano2023iris, bulekov2022morphuzz},~3)~the research gaps of integrating advanced techniques into traditional fuzzing workflows~\cite{Wang2019ASR, Zhang2018SurveyOD, Wang2020SoKTP, Zhu2022FuzzingAS, Mallissery2023DemystifyTF}, and 
4) the PUT \cite{Yun2022FuzzingOE, Eisele2022EmbeddedFA}. 
Here, we adopt the fourth dimension as the guiding taxonomy, as the first three dimensions primarily focus on the general study of fuzzing techniques and lack the specificity required for analyzing OSF. In contrast, the PUT-based classification provides~a~comprehensive framework for systematically reviewing the general workflow of OSF while focusing on the unique challenges posed by the different OS layers (\ie different PUTs).

\subsection{OS Layers}\label{Section3-2}


According to the PUT-based classification we employed, OSF can be categorized into four types, \ie kernel fuzzing, file system fuzzing, driver fuzzing, and hypervisor fuzzing. Due to the distinct differences in fuzzing approaches across these four OS layers, it is necessary to outline the primary functions of each OS layer and the primary tasks of their respective fuzzing approaches.


\begin{figure*}[!t]
        \centering
        \subfigure[Interfaces of Each OS Layer]{
                \includegraphics[width=0.38\linewidth]{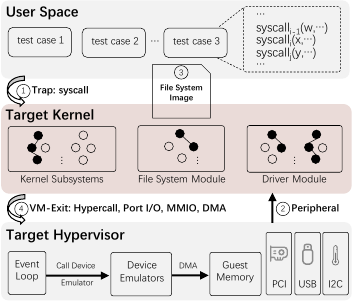}\label{OSFuzzing}}
        \hspace{0.03\textwidth}
        \subfigure[General Workflow of OSF]{
                \includegraphics[width=0.54\textwidth]{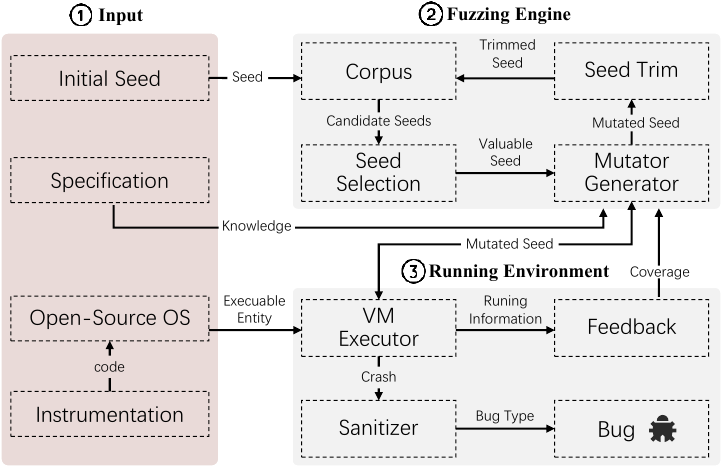}\label{OSF_WorkFlow}}
\vspace{-10pt}
        \caption{Interfaces of Each OS Layer (where \protect\circled{1}, \protect\circled{2}, \protect\circled{3} and \protect\circled{4} denote the fuzzing interfaces for Kernel, Driver, File System and Hypervisor respectively; and \protect\circled{1} is also the fuzzing interface for Driver and File System); and General Workflow of OSF (which has three modules, \ie input, fuzzing engine, and running environment)}
\end{figure*}


\textbf{Kernel}. Kernel is one of the most critical systems in an OS because it manages essential resources such as processes and memory for the entire system and provides a unified programming interface for user-space programs to interact with hardware resources. All other system types rely~on~and operate on top of the kernel. Consequently, vulnerabilities in the kernel can be maliciously exploited, potentially causing significant damage to the systems running on it. As illustrated in Figure \ref{OSFuzzing}, to perform kernel fuzzing, a fuzzer triggers the kernel's code paths by switching from user space~to kernel space through syscall \circled{1} (a.k.a. system call) \cite{Koopman1997ComparingOS}. Therefore, the core task of a kernel fuzzer is to generate various test cases (referred to as seeds) by combining syscalls~provided~by~the~kernel to continuously trigger the kernel's code paths. For example, the darker sequences in Figure \ref{OSFuzzing} denote branches covered by test cases, while the lighter circles denote code blocks yet to be covered.

\textbf{File System}. As a core system service within an OS, file system is essential for tasks such~as~reading, writing, managing, and scheduling files, as well as ensuring data consistency during system crashes. Most file systems, like ext4 \cite{Cao2007Ext4TN}, XFS \cite{XFS2018}, Btrfs \cite{Rodeh2013BTRFSTL} and F2FS \cite{Lee2015F2FSAN}, operate within~the~OS kernel. Therefore, approaches used in kernel fuzzing can be adapted for file system fuzzing. However, approaches based purely on syscalls often yield numerous invalid results because the system state is predominantly influenced by metadata in file system operations. In contrast, operations~on regular file data through syscalls like \texttt{read()} and \texttt{write()} contribute little to identifying file system vulnerabilities. Hence, an effective file system fuzzer typically combines sequences of file operation-related syscalls \circled{1} with images \circled{3} where metadata has been changed. As shown in Figure \ref{OSFuzzing}, the altered disk image is mounted to the file system's partition using privileged commands \cite{schumilo2017kafl, Hydra2020finding}, becoming the new target for fuzzing. Kernel-provided syscalls are then used to conduct read, write, management, and scheduling operations, facilitating a comprehensive fuzzing of the file system.

\textbf{Driver}. Drivers are responsible for communication and control between the user/kernel and hardware devices. They act as the bridge between hardware and OS, managing tasks such~as~device~initialization, data transfer, and interrupt handling, ensuring that user-space applications~can~interact with underlying hardware \cite{Chou2001AnES, Palix2011FaultsIL, Stoep2018AndroidSecurity}. In driver fuzzing, the primary objective is to uncover vulnerabilities in device drivers across the initialization, data communication, and control stages. Since drivers can receive operation requests from both user space and hardware devices, they expose a broader attack surface compared to the kernel or other kernel subsystems \cite{Beniamini2017Part1, Beniamini2017Part2, Chang2017, Davis2011, NohlLell2014}. As illustrated in Figure \ref{OSFuzzing}, the two main attack surfaces for driver fuzzing are syscalls \circled{1} and peripheral interfaces \circled{2}. Unlike kernel fuzzers, syscall-based driver fuzzers focus more on syscalls that directly operate on device files, \eg \texttt{read()}, \texttt{write()}, \texttt{seek()}, \texttt{ioctl()}, \etc

The peripheral interface can be exploited in two ways, \ie I/O interception and device configuration. I/O interception involves intercepting access to I/O objects (\eg DMA, MMIO, and Port I/O) to mutate I/O data, which is then redirected to the target to observe the driver's behavior.~Device configuration, on the other hand, simulates peripheral device behavior and injects the simulated data into the I/O channel. In addition, it is worth noting that drivers constitute the largest codebase among kernel subsystems, and exhibit significant variability due to implementations by different vendors. As a result, driver fuzzers often face larger challenges in achieving high coverage, generating diverse test cases, and ensuring fuzzing effectiveness given these characteristics.

\textbf{Hypervisor}. In environments with highly heterogeneous hardware, OSs inherently lack~the~capability to manage and schedule heterogeneous computing resources, such as those used in industries like railways, avionics, and automotive systems \cite{Cinque2021VirtualizingMS}. To enable the concurrent execution of multiple OSs while maintaining secure isolation between them in such heterogeneous resource environments \cite{RTCA_DO_178C, ISO_26262_2011}, hypervisors, a.k.a. Virtual Machine Monitors (VMMs), leverage virtualization~technique to partition hardware resources (\eg CPU, memory, and disk space) into multiple virtual partitions \cite{Popek1974FormalRF, Cinque2021VirtualizingMS, Cilardo2021VirtualizationOM}. When a guest OS, which runs within a Virtual Machine (VM), accesses or operates hardware devices, it triggers a VM-exit event that transfers these privileged operations~to~the~hypervisor for hardware emulation. Therefore, the objective of hypervisor fuzzing is to accurately emulate the behavior of the guest OS when accessing and operating these virtualized hardware components, which serves as the primary entry point for implementing hypervisor fuzzing trials. As illustrated in Figure \ref{OSFuzzing}, the main interfaces involved include I/O channels and hypercalls \circled{4}. 

To ease the understanding of how to perform hypervisor fuzzing using these interfaces,~we~introduce the technical background of hypervisors. Technically, hypervisors can be categorized into full virtualization and para-virtualization. In full virtualization, the guest OS is unaware~of~its~virtualized environment. Consequently, when the guest OS attempts to access physical hardware (\eg memory access), the hypervisor intercepts the request. At this point, the VM triggers a trap~and~enters~a~VM-exit state, with the hypervisor assuming a full control of the VM's operations. Subsequently, the hypervisor forwards the memory access request to the Device Emulator, which provides virtualization for Port I/O (input/output port instructions), MMIO (Memory-Mapped I/O for direct memory access), and DMA (for complex and large-scale data transfers). Finally, the Device Emulator returns the virtual memory access interface to the guest OS. In a fully virtualized environment, the I/O channel is more extensively utilized because this technique offers complete hardware emulation. However, this can also lead to more frequent communication overhead.


To facilitate and accelerate communication between the guest OS and the hypervisor, modern hypervisors often support hardware-accelerated virtualization technologies. By introducing specialized instructions (such as \texttt{vmcall}) to perform hypercalls, para-virtualization allows OSs to bypass the VM and communicate directly with the hypervisor. Therefore, in addition to the I/O channel, the hypercall interface can also be employed to implement hypervisor fuzzing trials. As a result, it is necessary to conduct specialized fuzzing of hypervisors to uncover vulnerabilities related to resource management and security isolation in generalized OSs.

\subsection{OSF Workflow}\label{Section3-3}

We summarize the general workflow of OSF in Figure~\ref{OSF_WorkFlow}, which consists of three key modules, \ie \textit{input}, \textit{fuzzing engine}, and \textit{running environment}. Each module is described in detail as follows.


\textbf{Input}. There are three types of inputs to be considered in the input module, \ie initial seed,~specification, and target OS. First, the initial seed is the raw material fed to the fuzzing engine, which~is initially deposited in a global corpus (\ie a repository for storing seed candidates) and subsequently used throughout the fuzzing workflow. Second, the specification, which describes the structure and syntax of the seed, is parsed by the fuzzing engine and used to generate new seeds or enhance~the quality of the seeds. It is important to note that the use of the specifications is not applicable~to~all scenarios; it mainly applies to highly structured seeds with documented descriptions of the seed structure \cite{Syzkaller, chen2020koobe, zhao2022statefuzz, sun2021healer, shen2021rtkaller, shen2022tardis, jeong2023segfuzz, xu2020krace, jeong2019razzer}. Third, the target OS requires to be instrumented to collect runtime information for feedback analysis and bug monitoring during fuzzing.

\textbf{Fuzzing Engine}. The main function of the fuzzing engine is to generate new seeds and wrap them into test cases for the input fed to the executor in the running environment. This module~go through a complete closed loop of seed selection from a corpus, seed mutation and/or generation,~and~seed~trim before storing to the corpus. Specifically, the corpus stores only seeds that have~been~verified~as ``high-quality'', where ``high-quality'' seeds can be defined as the seeds~that~have~triggered~bugs~\cite{zou2022syzscope, wu2018fuze,lin2022grebe,shen2021rtkaller}, the initial seed generated based on tailored rules \cite{shen2022drifuzz} or specifications \cite{Syzkaller},~or~the streamlined seeds after seed trim as they trigger new coverage \cite{pailoor2018moonshine, shen2021rtkaller}. Such a consideration stems from the empirical conclusion that {mutations} based on high-quality seeds usually have more opportunities to move the execution of the PUT closer to trigger bugs~\cite{you2017semfuzz, shen2022drifuzz, lin2022grebe, Unicorefuzz2019}.~However, even if the corpus always stores high-quality seeds, it is necessary to consider which~seeds should be prioritized in order to improve the efficiency of {seed mutation}. Therefore, seed selection~is~often adopted to prioritize the seeds in the corpus, and a computation method of seed priority is often devised to ensure that the selected seeds have more chances to reveal new bugs, \eg PageRank~\cite{lin2022grebe}, program analysis \cite{wu2018fuze, chen2022sfuzz}, evolutionary algorithm \cite{shen2022drifuzz, henderson2017vdf, xu2020krace}, empirical strategy \cite{xu2020krace, Syzkaller, you2017semfuzz, zou2022syzscope}, reinforcement learning \cite{wang2021syzvegas}, \etc Then, to continuously generate test cases~to~automate~the whole fuzzing process, {seed mutation and/or generation}  transform and/or generate the seeds in some way and wrap them into test cases that can be executed by the target OS. For example, seed mutation employs parameter-level bit flips or byte-level replacement to generate seeds based on the high-quality seeds in the corpus. Seed generation builds test case models from the specification~to generate new seeds. Finally, to further improve the speed performance of OSF, the generated~seed usually requires to be trimmed to reduce its mutation space as well as its execution time, \eg greedy algorithm~\cite{pailoor2018moonshine} and stepwise filtering \cite{schumilo2020hyper, henderson2017vdf}. The trimmed seeds are then stored into the corpus, and the process is repeated to automatically drive new and high-quality seed generation.

\textbf{Running Environment}. The running environment is responsible for feeding the test cases wrapped by the fuzzing engine into the executor. Unlike traditional fuzzing techniques, OSF~usually relies on full-virtual environment to avoid as much as possible the effects propagated by~OS~crashes. In contrast, traditional fuzzing techniques do not fatally affect the entire fuzzing process even if a crash occurs. Therefore, the running environment of OSF is often designed as full-virtual~environment that can be quickly recovered, allowing the entire system to be quickly restarted to a known, clean state in the event of a crash \cite{shen2022tardis, lin2022grebe, schumilo2021nyx, pan2021V-shuttle, zhao2022semantic, song2020agamotto}. Such a design allows~a~fuzzer to run thousands of test cases continuously and automatically in an isolated environment without worrying about the potential for lasting damage from individual test cases. Subsequent execution results fall into two categories. If the executor does not crash, the feedback will be collected~by the instrumented code, which is used to guide {seed mutation} of the fuzzing engine to steer~the~fuzzing towards uncovered program paths. If the executor triggers a crash, the instrumented code records the information and sends it to the bug analyzer. The bug analyzer identifies and categorizes errors, and generates detailed bug reports. Finally, if the fuzzing process is not yet finished, it should be restored to its pre-crash state, and then move on to explore other states of the target OS.

Although there are some similarities with traditional fuzzing in terms of the coarse-grained~fuzzing process, the technical challenges and implementation details of each step need to take into account the specificity and complexity of OS, which ultimately makes it show a big difference with traditional fuzzing in terms of the fine-grained {details}. Therefore, we will elaborate on the technical details and technical distinctions of each step illustrated in Figure \ref{OSF_WorkFlow} in Section~\ref{Section4} and \ref{Section5}.


\section{General Workflow Characteristics of OSF} \label{Section4}



We review the state-of-the-art OSF with respect to the key steps in the three core modules~of~the general fuzzing workflow, \ie input, fuzzing engine, and runtime environment, as shown in Figure~\ref{OSF_WorkFlow}. Our review results are reported in Table~\ref{tab1}, where ``-'' means that this dimension~is~not~relevant, or not clearly mentioned in the fuzzer; \textbf{ISG} denotes initial seed generation, \textbf{Inst.} denotes instrumentation, \textbf{SS}, \textbf{ST} and \textbf{SU} respectively denote seed selection, seed trim, and seed~update~(generation~or~mutation); \textbf{Feed.} denotes~feedback mechanism; and \textbf{BT} denotes supported bug types.

\begin{footnotesize}
        \begin{longtable}{m{2.1cm}m{1.3cm}m{0.6cm}m{1.5cm}m{0.9cm}m{0.8cm}m{0.9cm}m{1.5cm}m{0.8cm}}
        \caption{Characteristics of OS Fuzzers (Sorted by Publication Year)}\label{tab1} \vspace{-0.3cm}\\
      
        \toprule
        \multirow{2.5}{*}{\textbf{Fuzzer}} & \multirow{2.5}{*}{\textbf{OS Layer}} & \multicolumn{2}{c}{\textbf{Input}} & \multicolumn{3}{c}{\textbf{Fuzzing Engine}} & \multicolumn{2}{c}{\textbf{Run. Environment}}  \\
        \cmidrule(lr){3-4} \cmidrule(lr){5-7} \cmidrule(lr){8-9} & & \textbf{ISG} & \textbf{Inst.} & \textbf{SS} & \textbf{ST} & \textbf{SU} & \textbf{Feed.} & \textbf{BT} \\
        \midrule
        \endfirsthead
        
        \toprule
        \multirow{2.5}{*}{\textbf{Fuzzer}} & \multirow{2.5}{*}{\textbf{OS Layer}} & \multicolumn{2}{c}{\textbf{Input}} & \multicolumn{3}{c}{\textbf{Fuzzing Engine}} & \multicolumn{2}{c}{\textbf{Run. Environment}}  \\
        \cmidrule(lr){3-4} \cmidrule(lr){5-7} \cmidrule(lr){8-9} &  & \textbf{ISG} & \textbf{Inst.} & \textbf{SS} & \textbf{ST} & \textbf{SU} & \textbf{Feed.} & \textbf{BT} \\
        \midrule
        \endhead
        
        \bottomrule
        \endfoot

        Trinity\cite{Trinity} & Kernel & Spec. & - & - & - & Gen. & Code & -\\
        
        
        Syzkaller\cite{Syzkaller} & Kernel & Spec.' & Static & Feedback & DD & Mut. & Code & - \\

        TriforceAFL\cite{TriforceAFL} & Kernel & Pattern & Dynamic & Feedback & DD & Mut. & Code & M. \\
        
        DIFUZE\cite{2017DIFUZE} & Driver & Spec.' & Static & - & - & Gen. & - & M.,L.,O. \\
        
        VDF\cite{henderson2017vdf} & Hypervisor & Trace & Static & - & DD & Mut. & Code & M.,C.,L. \\
        
        kAFL\cite{schumilo2017kafl} & File System & - & Static+Dynamic & - & - & Mut. & Code & M. \\
        
        SemFuzz\cite{you2017semfuzz} & Kernel & PoC & Static & Distance & - & Mut. & Code & M. \\
        
        usb-fuzzer\cite{Syzkaller} & Driver & Spec.' & Static & - & - & Mut. & Code & M. \\
        
        Moonshine\cite{pailoor2018moonshine} & Kernel & Trace & Static & Feedback & - & Mut. & Code & M.,C. \\
        
        FUZE\cite{wu2018fuze} & Kernel & PoC & Static & - & - & Mut. & Code & M. \\
        
        Schwarz~et~al.\cite{schwarz2018automated} & Kernel & Trace & - & - & - & Gen. & Code & C. \\
        
        Razzer\cite{jeong2019razzer} & Kernel & Spec.' & Static & - & - & Mut. & Code & C. \\
        
        JANUS\cite{JANUS2019fuzzing} & File System & - & Static & - & - & Mut. & Code+Cust. & M.,L. \\
        
        SLAKE\cite{chen2019slake} & Kernel & PoC & Static & - & - & Mut. & - & M. \\
        
        Shi et al.\cite{shi2019industry} & Kernel & Spec.' & Static & - & - & Mut. & - & M.,C.,O. \\

        PeriScope\cite{Song2019PeriScopeAE} & Driver & Trace & Static & - & - & Mut. & Code & M. \\
        
        Unicorefuzz\cite{Unicorefuzz2019} & Kernel & Pattern & Static & - & - & Mut. & Code & M. \\
        
        KOOBE\cite{chen2020koobe} & Kernel & PoC & Dynamic & - & - & Mut. & Code+Cust. & M. \\
        
        Krace\cite{xu2020krace} & File System & Spec.' & Static & Minimum & - & Mut. & Code+Thread & C. \\
        
        Hyper-Cube\cite{schumilo2020hyper} & Hypervisor & Pattern & - & - & DD & - & - & M.,C.,L.,O. \\
        
        Hydra\cite{Hydra2020finding} & File System & - & Static & - & - & Mut. & Code+Cust. & M.,L.  \\
        
        USBFuzz\cite{peng2020usbfuzz} & Driver & Pattern & Static & - & - & Mut. & Code & M. \\
        
        Agamotto\cite{song2020agamotto} & Driver & Trace & Static & - & - & Mut. & Code & - \\

        Ex-vivo\cite{Pustogarov2020ExvivoDA} & Driver & Trace & Static & - & - & Mut. & Code & M. \\
        
        HFL\cite{kim2020hfl} & Kernel & Spec.' & Static & - & - & Mut. & Code & M. \\
        
        X-AFL\cite{liang2020xafl} & Kernel & Trace & Static & - & - & Mut. & Code & M. \\
        
        HEALER\cite{sun2021healer} & Kernel & Spec.' & Static & - & - & Gen.+Mut. & Code & M.,C.,L. \\
        
        Rtkaller\cite{shen2021rtkaller} & Kernel & Spec.' & Static & - & - & Mut. & Code & M.,C.  \\
        
        NYX\cite{schumilo2021nyx} & Hypervisor & Pattern & Dynamic & - & - & Mut. & Code & M. \\
        
        V-Shuttle\cite{pan2021V-shuttle} & Hypervisor & Trace & Static & - & - & Mut. & Code & M.,L. \\
        
        BSOD\cite{maier2021bsod} & Driver & Pattern & Dynamic & - & - & Mut. & Code & M. \\
        
        SyzVegas\cite{wang2021syzvegas} & Kernel & Spec.' & Static & Feedback & - & Mut. & Code & M. \\
        
        StateFuzz\cite{zhao2022statefuzz} & Driver & Spec.' & Static & Similarity & - & Mut. & Code+Cust. & M. \\
        
        GREBE\cite{lin2022grebe} & Kernel & PoC & Static & Minimum & - & Mut. & Cust. & M.,O. \\
        
        MundoFuzz\cite{myung2022mundofuzz} & Hypervisor & Trace & Static & - & - & Mut. & Code & M.,L. \\
        
        Morphuzz\cite{bulekov2022morphuzz} & Hypervisor & Trace & Static & - & - & Mut. & Code & C.,L. \\
        
        CONZZER\cite{CONZZER2022context} & File System & Pattern & Static & Feedback & - & Mut. & Code+Thread & C. \\
        
        
        Dr.Fuzz\cite{zhao2022semantic} & Driver & Spec.' & Static+Dynamic & - & - & Mut. & Code+Cust. & M.,O. \\
        
        PrIntFuzz\cite{ma2022printfuzz} & Driver & Spec.' & Static+Dynamic & - & - & Gen.+Mut. & Code & M.,C. \\
        
        DriFuzz\cite{shen2022drifuzz} & Driver & Trace & Static & - & - & Gen.+Mut. & Code & M. \\
        
        Hao~et~al.\cite{Hao2022DemystifyingTD} & Kernel & - & Static & - & - & - & - & - \\
        
        KSG\cite{sun2022ksg} & Kernel & Spec.' & Static & - & - & - & - & M.,C.,L. \\
        
        SyzScope\cite{zou2022syzscope} & Kernel & PoC & Static & - & - & Mut. & - & M. \\
        
        Tardis\cite{shen2022tardis} & Kernel & Spec. & Static & - & - & Mut. & Code & M.,L. \\
        
        SegFuzz\cite{jeong2023segfuzz} & Kernel & Spec.' & Static & - & - & Mut. & Code+Thread & M.,C.,O.  \\
        
        IRIS\cite{cesarano2023iris} & Hypervisor & Trace & Static & - & - & Mut. & Code & M. \\
        
        FUZZNG\cite{bulekov2023FUZZNG} & Kernel & Pattern & Static & - & - & Mut. & Code & M.  \\
        
        ACTOR\cite{fleischer2023actor} & Kernel & Trace & Static & - & - & Gen. & Code+Cust. & M.,L. \\
        
        SyzDescribe\cite{hao2023syzdescribe} & Driver & Spec.' & - & - & - & - & - & - \\
        
        DEVFUZZ\cite{wu2023devfuzz} & Driver & Pattern & Static+Dynamic & - & - & Mut. & Code+Cust. & M.,L.,O. \\
        
        SyzDirect\cite{tan2023syzdirect} & Kernel & Spec.' & Static & Distance & - & Mut. & Code & - \\
        
        ReUSB\cite{Jang2023ReUSB} & Driver & Trace & Dynamic & - & Protocol & Mut. & Code & M. \\
        
        DDRace\cite{Yuan2023DDRace} & Driver & Spec.' & Static & Feedback & - & Mut. & Code+Thread & M.,C. \\
        
        VD-Guard\cite{Liu2023VDGuard} & Hypervisor & Pattern & Static & - & - & Mut. & Code+Cust. & M.,L. \\
        ViDeZZo\cite{Liu2023ViDeZZoDV} & Hypervisor & Trace & Static & - & - & Gen.+Mut. & Code & M. \\
        
        Lfuzz\cite{Liu2023LFuzz} & File System & - & Static & - & - & Mut. & Code & M.,L. \\

        KernelGPT\cite{yang2023kernelgpt} & Kernel & Spec.' & Static & - & - & Mut. & Code & M.,O. \\

        BRF \cite{Hung2024BRFFT} & Kernel & Spec.' & Static & - & - & Gen.+Mut. & Code & M.,C.  \\

        MOCK \cite{Xu2024MOCKOK} & Kernel & Spec.' & Static & - & - & Gen.+Mut. & Code & M.,C. \\

        SATURN\cite{Xu2024Saturn} & Driver & Spec.' & Static & Similarity & - & Gen.+Mut. & Code & M.,C.,I.  \\

        Syzgen++\cite{Chen2024SyzGen++} & Driver & Spec.' & Static & - & - & - & Code & M.,C.,I. \\

        VIRTFUZZ\cite{Huster2024ToBoldly} & Driver & Trace & Static & - & - & Mut. & Code & M.,L. \\

        Monarch\cite{TAO2024MONARCH} & File System & Spec' & Static & - & - & Mut. & Code & M.,C. \\

        HYPERPILL\cite{Bulekov2024HYPERPILLFF} & Hypervisor & Pattern & Static & - & - & Mut. & Code & M. \\

        
        \end{longtable}
\end{footnotesize}
        

\subsection{Initial Seed Generation}\label{sec4.1}

Initial seed generation focuses on how to generate seeds at the beginning of OSF, which can~be~categorized into four types, \ie pattern-, specification-, trace-, and PoC-based approaches. 

\subsubsection{Pattern-Based}




Pattern-based approaches construct initial bytecode sequences based~on~predefined input structure templates, such as field lengths and bit widths. These approaches are also commonly employed in userland fuzzing, notably utilized by open-source tools such as AFL~\cite{AFL} and libFuzzer~\cite{libfuzzer} 
for constructing initial seeds. These tools were not specifically designed~for~OSF. However, because specialized fuzzing tools for OSF emerged relatively late, early research on OSF often relied on these general-purpose fuzzing tools to generate initial seeds.

In practice, pattern-based approaches typically require testers to explicitly define the structure of input data in the harness module, including details such as field lengths and bit widths.~The~harness then extracts corresponding initial seeds from randomly generated bytecode sequences based on these defined structures. Such extraction heavily relies~on testers' deep understanding of the input structure—for instance, accurately specifying the bit widths of specific fields within byte sequences (\eg vendor information fields required for driver validation). Hence, pattern-based approaches tend to incur high manual effort, as testers must explicitly specify the length and structure of each field within the harness to ensure the validity of initial seeds and accuracy of the fuzzing process.

For example, driver fuzzers USBFuzz \cite{peng2020usbfuzz}, DEVFUZZ \cite{wu2023devfuzz} and BSOD \cite{maier2021bsod} extend AFL~to~generate device~inputs (\eg vendor IDs and communication data). These fuzzers generate initial~seeds based on interaction patterns between drivers and devices, which are passed to the respective~drivers to simulate realistic hardware behavior in response to driver read requests. Since~AFL cannot directly~generate syscall sequences, kernel fuzzer TriforceAFL~\cite{TriforceAFL} and file system fuzzer~CONZZER~\cite{CONZZER2022context} extend AFL by mapping the generated bytecode to corresponding syscalls, thereby generating~initial seeds. To avoid manually writing extensive and complex parameter descriptions for each syscall, FUZZNG~\cite{bulekov2023FUZZNG} sets hooks at key kernel APIs to capture the ``specific address and byte count''.~Then it uses libFuzzer's random bytecode to dynamically~populate user~buffers or file descriptors, constructing valid parameters for syscalls. 
Several hypervisor~fuzzers~\cite{schumilo2020hyper,Bulekov2024HYPERPILLFF,schumilo2021nyx,Liu2023VDGuard}~rely~on manually-defined interaction patterns between virtual devices and I/O register interfaces (\eg~semantics of specific bit widths), and use tools like AFL and libFuzzer to generate initial seeds.

\subsubsection{Specification-Based}

Specification-based approaches construct a dedicated input specification for the PUT to guide the generation of initial seeds, ensuring that they meet basic syntactic rules. As data structures for I/O communications are implemented by third-party vendors~and~are~commercially protected, existing fuzzers only construct specifications for syscalls. In that sense, specification-based approaches are primarily adopted in scenarios where syscalls are used as seeds~\cite{pailoor2018moonshine, jeong2023segfuzz,xu2020krace}. 


Early popular kernel fuzzers like Trinity \cite{Trinity} use hard-coded rules as the specification to generate initial syscall sequences. For example, these rules specify a list of file descriptors, and annotate arguments with valid or near-valid data types and values. It is suitable for scenarios focusing solely on random syscalls generation \cite{xu2020krace}, but it is less adaptable to kernel evolution. For embedded~OS fuzzing, Tardis~\cite{shen2022tardis} manually writes syscall descriptions for core subsystems of the kernel (\eg~task and memory management) as well as a few commonly used peripherals, explicitly annotating the dependencies between syscalls. Upon startup, Tardis parses these descriptions into internal data structures, fills default parameters for each syscall, and then assembles multiple syscalls into a minimal feasible invocation sequence based on dependency~relations, which serves as the initial seed. Due to the large human effort, only a small number of syscall descriptions are supported. 


To cover diverse and accurate syscalls and ease their specification, Syzkaller \cite{Syzkaller} uses a structured description language, called syzlang \cite{syzlang}, to support more comprehensive syscall declarations,~and thus provides more semantic information during initial seed generation. Since its introduction, various fuzzers directly use Syzkaller as the foundational infrastructure for initial seed generation, which are marked as \textit{Spec.'} in Table \ref{tab1}.
To improve the quality of the initial seed generated~by~Syzkaller, many fuzzers~\cite{2017DIFUZE,kim2020hfl,ma2022printfuzz,sun2022ksg,hao2023syzdescribe,tan2023syzdirect,Yuan2023DDRace,Hung2024BRFFT,Xu2024MOCKOK,Xu2024Saturn,Chen2024SyzGen++} extract entry points and infer dependencies of syscalls using program analysis techniques, manual analysis based on domain knowledge, and neural network models. More recently, KernelGPT~\cite{yang2023kernelgpt} employs LLM to construct syzlang specification automatically, enabling Syzkaller to support newly merged code in the kernel.

\subsubsection{Trace-Based}

To avoid the complex construction of specifications, trace-based approaches use real applications or devices as trigger engines to create initial seed sequences based on intercepted executions. Some fuzzers intercept syscall sequences \cite{pailoor2018moonshine,schwarz2018automated,liang2020xafl,fleischer2023actor,Jang2023ReUSB}, I/O communication~\cite{song2020agamotto,Pustogarov2020ExvivoDA,pan2021V-shuttle,myung2022mundofuzz,bulekov2022morphuzz,shen2022drifuzz,Song2019PeriScopeAE,Liu2023ViDeZZoDV,Huster2024ToBoldly}, or driver operations \cite{henderson2017vdf,cesarano2023iris,Jang2023ReUSB} to generate initial seeds. These approaches ensure that the initial seeds are real and effective, enhancing the fuzzer's flexibility and portability. However, while trace-based approaches alleviate the need for manual specification, the absence of structural usage information results in structure-unaware mutations, undermining the fuzzer’s capability to generate valid seeds and achieve deep fuzzing coverage. Trace-based initial seed generation approaches can be implemented through software, hardware instrumentation, or third-party tools like STRACE \cite{Strace}, Wireshark \cite{Wireshark}, USBMON \cite{usbmon}, and Flush+Reload \cite{Yarom2014FLUSHRELOADAH, 2011IS}.

\subsubsection{PoC-Based}


PoC-based approaches generate initial seeds by taking PoCs or vulnerability~reports as input, where the PoCs or vulnerability reports can be acquired from CVE~\cite{cve}, Linux~git logs~\cite{linuxkernel}, Syzbot~\cite{Syzbot} and vulnerability descriptions published in forums and blogs. Fuzzers that employ such approaches are primarily designed to generate vulnerability exploits or assess vulnerability severity rather than identifying unknown vulnerabilities.

For example, SemFuzz~\cite{you2017semfuzz} utilizes natural language processing to extract key syscalls and~their parameters from CVE reports and patches as initial seeds. For unknown parameters, SemFuzz builds a syscall prototype knowledge base and a call dependency graph based on the Linux Programmer Manual (LPM), 
allowing it to fill in the values to the unknown parameters and generate~syscall sequences with complete dependencies. As Syzbot records an increasing number of PoCs~for~vulnerabilities, tools such as FUZE~\cite{wu2018fuze}, KOOBE~\cite{chen2020koobe}, SyzScope~\cite{zou2022syzscope}, GREBE~\cite{lin2022grebe}, and SLAKE~\cite{chen2019slake} 
directly collect syscall sequences from Syzbot's PoCs, utilizing these sequences as initial seeds. 
Note that the primary goal of these tools is to evaluate whether the seeds triggering these vulnerabilities miss broader exploit paths. Therefore, the seed information from these PoCs (including syscall sequences and vulnerability stack traces) typically serves as triggers, which are then analyzed~using program analysis techniques, \eg taint analysis \cite{zou2022syzscope, lin2022grebe}, static analysis \cite{chen2019slake},~and~symbolic execution \cite{wu2018fuze, chen2020koobe}, to investigate the vulnerability objects and their potential exploit paths.

\subsection{Instrumentation}\label{subsection4.2}

Instrumentation inserts probes into a program to collect runtime information from the PUT~while~preserving its original functionality and logical structure \cite{Instrumentation_Huang1978, ibm2020}. By analyzing the collected runtime~information, insights into the PUT's control and data flow can be acquired. Consequently,~this~allows for the calculation of coverage metrics and~the~monitoring of bugs. Such data then guides~the~fuzzing process towards more sensitive and potentially critical code paths. Unlike userland fuzzing, instrumentation in kernel space using built-in compiler tools such as gcov~\cite{gcov} and ASan~\cite{ASAN}~is~impractical due to the large scale and complexity of the modern OSs. Consequently, researchers~have~developed two types (\ie static and dynamic) of instrumentation approaches for OSs.

\subsubsection{Static Instrumentation}

Static instrumentation typically involves inserting probe programs into the source code or intermediate representation code. There is a rich set of tools for static~instrumentation for OSs, such as kcov \cite{vyukov2018kcov} and KASAN \cite{kernel2018kasan}. These tools are well-suited for OSF~due to their advantages of being easy to use, having relatively low overhead, being customizable,~and providing strong support for Unix-like OSs. For instance, Syzkaller \cite{Syzkaller} and Syzkaller-based works (\eg \cite{schumilo2017kafl, pailoor2018moonshine}) compile kcov and KASAN into the kernel to guide OSF based on collected code~coverage and detect memory-related bugs. kcov and KASAN were introduced in Linux versions~4.6~and~4.0, respectively. For older versions of Linux, SemFuzz \cite{you2017semfuzz} implemented the porting of these tools.

However, these built-in instrumentation tools are not always effective in certain scenarios, such as exploring the program state space, handling thread interleaving for concurrency errors,~and~performing directed fuzzing. Some studies have addressed these challenges by extending existing~compiler tools to implement fine-grained and targeted instrumentation. For example, coverage-guided~fuzzers often discard test cases, which are actually useful for exploring potential program states, if these test cases do not trigger new code coverage. To address this problem, StateFuzz \cite{zhao2022statefuzz} introduces a new feedback metric called state coverage, and uses the static instrumentation tool LLVM SanCov~\cite{llvm2024} to collect program state information, and thus fine-tunes the execution direction of coverage-guided fuzzers. In the context of concurrency error detection, fuzzers such as DDRace \cite{Yuan2023DDRace}, Krace \cite{xu2020krace}, SegFuzz \cite{jeong2023segfuzz}, CONZZER \cite{CONZZER2022context} and Razzer~\cite{jeong2019razzer} design customized thread coverage metrics to guide data race detection through the LLVM instrumentation suite. For directed fuzzing, GREBE \cite{lin2022grebe} uses LLVM Analysis and Pass to track taint propagation paths in the program to identify critical objects. Tardis \cite{shen2022tardis} addresses the unavailability of kcov in embedded OSs (\eg FreeRTOS) by leveraging Clang's SanitizerCoverage~\cite{llvm_sancov}, and proposes an efficient coverage collection callback.


\subsubsection{Dynamic Instrumentation}

Dynamic instrumentation, or dynamic binary instrumentation~(DBI), happens while the PUT is running. Although DBI has a higher runtime overhead compared~to~static instrumentation, its flexibility makes it useful for OSs without built-in instrumentation tools or for earlier versions of OSs. DBI can be categorized into hardware- and software-assisted techniques.

\textbf{Hardware-Assisted Technique.}
Hardware-assisted DBI (\eg Intel Processor Trace \cite{kleen2015,simplept} and ARM CoreSight \cite{coresight2017}) leverages special CPU features to trace the execution and branch information of the PUT. Hardware-assisted DBI records the execution paths rather than code coverage, but the detailed execution path information can be used to infer runtime coverage. Intel and ARM offer similar features. The former is suitable for testing kernels, drivers, file systems, and hypervisors in virtualized environments, while the latter is mainly used in embedded systems and mobile devices. Although ARM CPUs may support CoreSight, this feature is not mandatory in the ARM~specification. In practice, many commercial Android devices either use ARM CPUs without this feature or disable it due to power, cost, or security constraints, rendering hardware-assisted tracing impractical \cite{Chizpurfle2019}.

One important motivation for adopting hardware-assisted instrumentation is its ability to address limitations in traditional coverage collection mechanisms. For example, in driver fuzzing,~one~common challenge is to collect coverage information during the early initialization phase of device~drivers. At this phase, the \texttt{debugfs} interface, through which kcov exposes coverage data, is not~yet~available. 
To overcome this problem, Dr.Fuzz \cite{zhao2022semantic}, PrIntFuzz \cite{ma2022printfuzz}, DEVFUZZ \cite{wu2023devfuzz} and kAFL \cite{schumilo2017kafl} leverage Intel Processor Trace~to~track~the~execution flow during the driver initialization phase. Then, they switch back to kcov to obtain~precise coverage information. They effectively enhance~the depth of driver fuzzing through combining the strengths of static and dynamic instrumentation.

\textbf{Software-Assisted Technique.}
Software-assisted DBI refers to the dynamic injection of binary instructions during program execution using software breakpoints or binary rewriting techniques. This technique is more flexible and generally applicable to devices without specific hardware~support. 
As software-assisted DBI typically incurs a higher runtime overhead, this technique should be treated with a grain of salt due to concerns about throughput and flexibility.

Software-assisted DBI is commonly realized through INT 3 software breakpoints \cite{2006Intel6} and~QEMU \cite{bellard2005qemu}. INT 3 is an instruction in the x86 and x86-64 architectures that is used to trigger debug~interrupts. Specifically, it obtains the execution flow information of the PUT by replacing the basic block jump instruction or the conditional control flow instruction after disassembling the binary PUT into a control flow graph. Eventually, the code coverage calculation for the binary PUT can be realized by integrating a modified Syzkaller's kcov module or AFL's coverage calculation~module.~For~example, BSOD \cite{maier2021bsod} collects program control flow information by pausing the VM and~replacing the first byte of the control flow instruction with the \texttt{0xcc} instruction (\ie INT 3). 

QEMU is a full-system emulator that supports multiple instruction set architectures. At its~core~lies the Tiny Code Generator (TCG), which dynamically translates guest CPU instructions into~an~intermediate representation (TCG IR) at runtime, and subsequently performs just-in-time~compilation into native instructions for the host CPU, which are then cached as translation blocks. Building on this mechanism, TriforceAFL~\cite{TriforceAFL} hooks into QEMU's TCG, and inserts probes at the entry of each translation block to log control flow transitions and determine whether an edge has been covered, thereby enabling coverage-guided fuzzing. Moreover, TriforceAFL hooks the symbol addresses of critical kernel error-handling functions, such as \texttt{panic()} (invoked upon unrecoverable fatal errors) and \texttt{log\_store()} (logging diagnostic messages). When the execution path reaches either of these functions, the current input is marked as triggering a crash and reported back to TriforceAFL. 

\subsection{Seed Selection}
\label{Section4.3}


Seed selection aims to prioritize relatively ``valuable'' seeds from the corpus for subsequent mutation, which is a particularly important step in determining the direction of seed evolution. Therefore, various seed selection strategies have been designed and adopted in OSF tools.

\subsubsection{Minimum Frequency.}
Past practices have demonstrated that executing rarer code paths~helps test corner situations and makes it easier to expose bugs \cite{Bx00F6hme2016CoverageBasedGF, Lemieux2017FairFuzzAT}. Therefore, the minimum frequency strategy is to select the least frequently used seeds for the next round of fuzzing. Although~this strategy is not optimal, its simplicity and effectiveness can achieve fuzzing objectives. For example, Krace~\cite{xu2020krace} selects the two least used seeds each time and merges them into two different~threads~to explore data race bugs, and GREBE \cite{lin2022grebe} employs PageRank \cite{Brin1998PageRank} to eliminate popular kernel objects. 
However, due to significant differences in code size across OS subsystems, high-frequency seeds~are not necessarily inefficient. For example, the \texttt{autofs} subsystem has only four bugs fixed, while the larger \texttt{bcachef} subsystem has 303 fixed bugs \cite{Syzbot}, many of which were triggered by high-frequency syscalls. Relying solely on minimum frequency might overlook valuable seeds that, despite being frequently used, have already constructed complex conditions required to trigger vulnerabilities.


\subsubsection{Feedback-Guided.} \label{feedback-guided seed selection}
To improve the fuzzing coverage, a more straightforward strategy~is~to~select seeds that contribute to overall code coverage. For example, TriforceAFL \cite{TriforceAFL} builds on~AFL's~core mechanisms, prioritizing seeds via a comprehensive scoring mechanism based on metrics such as new coverage, execution speed, path depth, and novelty, whereas Syzkaller \cite{Syzkaller} selects seeds via a weighted random strategy, where seeds with higher coverage have a greater probability of being selected. It is worth mentioning that in Table \ref{tab1}, the fuzzers based on syzkaller (those with \textit{Spec.'} in the ISG column) inherit Syzkaller's seed selection strategy.
Moonshine \cite{pailoor2018moonshine} prioritizes seed based on code coverage in descending order, ensuring that each selected seed is most beneficial~for improving global coverage.
SyzVegas~\cite{wang2021syzvegas} adopts the EXP3 algorithm \cite{Blondel2008FastUO} 
to dynamically~adjust the seed selection probability, prioritizing those that yield higher coverage and incur lower~execution time costs.
Beyond code coverage feedback, some directed fuzzers~\cite{CONZZER2022context,Yuan2023DDRace} customize new coverage feedback to guide seed selection. For example, CONZZER \cite{CONZZER2022context} and DDRace \cite{Yuan2023DDRace} customize thread interleaving coverage to prioritize seeds for two threads that are more likely to trigger data race bugs.
Feedback-guided fuzzing has achieved significant success, therefore most existing works continue to follow Syzkaller’s seed selection strategy. New seed selection strategies are typically introduced only for specific tasks, such as targeting certain bug types or specific kernel modules.

\subsubsection{Shortest Distance.}
Another seed selection strategy specifically designed for directed fuzzing~is the shortest distance strategy. It constructs a call graph for kernel objects, and then uses the distance to the target site as the primary criterion for seed selection. For example, SemFuzz \cite{you2017semfuzz} constructs the call graph by modifying GCC to collect call information during kernel compilation, and~uses the inverse of the distance from each candidate input's reachable functions to the vulnerable function as the priority of seed selection. SyzDirect \cite{tan2023syzdirect} statically identifies target points (\eg vulnerability sites, and patch locations), and constructs a call graph between syscalls and these target points. It prioritizes seeds whose reachable paths have the shortest distance to the target points.

The shortest distance to the target site reflects the relative importance of a seed in the field~of directed fuzzing. However, seeds selected based on this metric may deviate from their original~evolution path or fall into local mutations after random changes, causing subsequent seeds to consistently fail to reach the target site \cite{Huang2022BEACONDG, Zong2020FuzzGuardFO}. An effective approach, adopted by SyzDirect \cite{tan2023syzdirect}, to address this issue is to construct~reachability templates for target sites, such as input dependencies and~key parameter constraints,~to~correct the deviation in seed evolution. 
SyzDirect's capability to build accurate templates benefits from the fact that the fuzzing targets are predefined patch locations, often accompanied by known PoCs. However, when the target shifts to a subsystem, module, or a class of vulnerabilities without readily available reachability templates, constructing accurate templates from scratch to guide seed selection and correction remains a significant challenge.


\subsubsection{Similarity Clustering.}
To avoid selecting similar seeds for mutation, a seed selection~strategy based on similarity clustering groups seeds based on their shared features, and assigns selection probability to each group. For example, StateFuzz \cite{zhao2022statefuzz} clusters seeds and ensures equal~probability for selecting seeds across different clusters. In collaborative fuzzing scenarios, SATURN~\cite{Xu2024Saturn}~categorizes the corpus by device functionality (\eg printers, keyboards, and storage devices),~and~dynamically selects seeds that enable host-device interaction based on the currently attached device~type, preventing potential seed interaction conflicts during collaborative fuzzing. 
Existing approaches select seeds with equal probability from different clusters to prevent over-exploration of similar seeds. However, placing too much emphasis on equality might lead to overlooking more valuable seeds, such as those capable of triggering deeper execution paths or those that have already constructed complex preconditions necessary for exposing vulnerabilities.

\subsection{Seed Trim}

Seed trim, also known as seed minimization, is to remove part of a seed that does not contribute~to~the fuzzing goal (\eg coverage improvement, and crash reproduction). 
It is considered as a critical~step~in ensuring efficiency, as redundant part of a seed wastes computational resources. 

\subsubsection{Delta Debugging-Based.}
The delta debugging (DD)-based minimization strategy aims~to~iteratively trim parts of a seed that do not contribute to coverage improvement or crash triggering.~In kernel fuzzing, both TriforceAFL \cite{TriforceAFL} and Syzkaller \cite{Syzkaller} minimize seeds based on coverage feedback, but their strategies differ. Specifically, TriforceAFL periodically compresses the entire corpus~by~identifying and removing seeds with redundant edge coverage, ultimately extracting a minimal subset~of seeds that together preserve full edge coverage. In contrast, Syzkaller maintains a set of covered edges, and only adds seeds that trigger new edges. Hence, it only needs to perform stepwise minimization on individual seeds before inserting them into the corpus. Particularly, Syzkaller~iteratively removes individual syscalls to extract the shortest syscall sequence, and then recursively optimizes syscall parameters, such as setting pointers to \texttt{NULL}, randomly deleting array elements, reducing buffer sizes, or shortening file names. As described in Sec.~\ref{feedback-guided seed selection}, the \textit{Spec.'} entries in the \textbf{ISG} column in Table \ref{tab1} also inherit Syzkaller’s seed minimization strategy. In hypervisor fuzzing, both VDF \cite{henderson2017vdf} and Hyper-Cube \cite{schumilo2020hyper} iteratively remove I/O messages from the seed that trigger crashes or hangs, until further reduction causes the failure to disappear.

Coverage-guided seed minimization effectively retains a minimal corpus required to achieve~the maximal coverage, thereby improving fuzzing efficiency. However, over-aggressive trimming~might also remove critical preconditions necessary for triggering certain bugs, 
potentially missing the opportunity to expose them. Therefore, current trim strategies still face challenges in balancing seed minimization and the preservation of critical preconditions.

\subsubsection{Protocol-Aware.}
Protocol-aware minimization strategy aims to remove input fields that do not affect the target device's semantic state or protocol handling behavior. ReUSB \cite{Jang2023ReUSB} utilizes~a~USB protocol parser to identify critical fields in the input packet, and retains only those that are actually parsed by the device driver and affect the device state, while removing semantically irrelevant~or redundant parts. For example, a raw SET\_REPORT request (used to send class-specific report data to a USB device) contains a 64-byte payload, 
of which only part is parsed by the device driver in the \texttt{usb\_set\_report} function (which processes the report request and updates internal state). The remaining bytes neither participate in protocol state transitions nor affect control-flow coverage. ReUSB identifies this semantic boundary and trims the payload, reducing unnecessary operations (such as memory copy or checksum) that do not cause state transitions.

Compared to coverage-based minimization strategies, this strategy is better suited for fuzzing state-machine-based device drivers. By leveraging protocol awareness, it identifies and removes input fields that do not contribute to state transitions, thereby improving the semantic validity~and execution efficiency of test inputs. It is worth mentioning that this strategy heavily relies on~the accurate modeling of protocol formats and precise analysis of driver behavior, which may limit its applicability to interfaces with complex or non-standard protocols.

\subsection{Seed Update}\label{Update}

Seed generation and mutation are two common methods for seed update~\cite{Mans2018TheAS,Sutton2007FuzzingBF}. In OSF,~seed~update strategies focus more on mutation or a combination of both to minimize the randomness of OSF, enhancing the effectiveness in terms of code coverage and bug-finding capability.

\subsubsection{Generation}


Generation-based methods generate new seeds using only the~initial seed generation techniques (see Sec \ref{sec4.1}) throughout the entire OSF campaign. Generation-based OSF work~is relatively scarce, appearing mainly in early OSF studies that fail to generate syntactically valid~seeds (\eg \cite{Trinity,2017DIFUZE}) and scenarios that heavily rely on extremely strict seed structures to trigger vulnerabilities (\eg memory double-fetch vulnerabilities triggered by manipulating complex structure~objects \cite{schwarz2018automated} and memory corruption vulnerabilities exposed only by specific memory actions \cite{fleischer2023actor}).



\subsubsection{Mutation}
Mutation-based methods drive the evolution of old seeds through mutation strategies (i.e., mutators for each seed type). Typical types include syscall, argument, and thread.

\textbf{Syscall.}
Syzkaller \cite{Syzkaller} is a classic mutation-based fuzzer, and syscall mutators have been designed, including adding, removing, and splicing. Adding is to insert a new syscall at the end of the existing syscall sequence; removing is to randomly delete a syscall from the sequence; and splicing is to randomly select a syscall sequence from seed corpus and splice it to the current sequence, with the splicing point also chosen randomly. Fuzzers using Syzkaller as the mutation engine (fuzzers listed in the \textbf{ISG} column as \textit{Spec.'} in Table \ref{tab1}) typically use these mutators for syscall mutation~or~adjust~the probability or usage of these mutators with novel feedback mechanisms (see Sec \ref{sec4.6}).

\textbf{Argument.}
Argument mutators focus on mutating the program inputs, including specific~arguments of syscalls (such as integers and flags), certain~fields in file system metadata, and even command bytes in I/O messages of drivers and hypervisors. Basic fuzzing engines like Syzkaller~\cite{Syzkaller}, AFL \cite{AFL}, and libFuzzer \cite{libfuzzer} have implemented various mutators for this purpose, including bit flipping, integer addition and subtraction, value replacement, and insertion of random bytes.
Most existing OSF works directly adopt the mutators implemented in basic fuzzing engines, though with different focuses in their application. For example, for kernel fuzzing, most works, such as those based on Syzkaller (fuzzers listed in the \textbf{ISG} column as \textit{Spec.'} in Table \ref{tab1}), apply bit flipping, integer addition and subtraction, value replacement, and random byte insertion to mutate syscall argument~values. For file system fuzzing, kAFL \cite{schumilo2017kafl} utilizes AFL to perform bit flipping on binary image files; and fuzzers like \cite{JANUS2019fuzzing,  Liu2023LFuzz} focus more on bit flipping within metadata blocks. For driver and hypervisor fuzzing, mutation is typically based on AFL or libFuzzer. Some efforts target mutations of data~flows that comply with device protocols (\eg USB and VirtIO) \cite{peng2020usbfuzz, Huster2024ToBoldly, Jang2023ReUSB} or pointer arguments in \texttt{ioctl} syscall \cite{Pustogarov2020ExvivoDA, maier2021bsod}; other works \cite{Song2019PeriScopeAE, wu2023devfuzz, Liu2023VDGuard, pan2021V-shuttle, henderson2017vdf, Bulekov2024HYPERPILLFF} apply mutators directly to I/O data streams of low-level hardware channels (such as MMIO and DMA).


\textbf{Thread.}
Thread mutators focus on how to perturb the scheduling of threads (each consisting of a complete syscall sequence) to expose concurrency bugs. There exist various methods for perturbing threads, such as injecting breakpoints or introducing thread delays; however, a more critical step before perturbation is to identify the specific kernel code regions where such perturbation should be applied, rather than perturbing arbitrarily. 

Razzer \cite{jeong2019razzer} splits the forward subsequences of two syscalls that may access the same shared~memory within a syscall sequence, treats them as two separate threads, binds them to different virtual CPUs (vCPUs), and mutates the parameters of hypercalls (a function call provided by QEMU \cite{bellard2005qemu} to control vCPU scheduling order) to alter the execution order of the threads. Krace \cite{xu2020krace}~injects~a large number of hooks into functions related to memory allocation and deallocation (\eg kmalloc and kfree). When a syscall triggers one of these hooked functions, it is paused and randomly~injected with microsecond-level delays to perturb thread interleaving. CONZZER \cite{CONZZER2022context} assumes that if a pair of concurrent functions has a potential data race, then their neighboring functions (\ie functions adjacent in the call stack) may also exhibit concurrency issues. To thoroughly test such neighboring concurrency vulnerabilities, CONZZER first executes multi-threaded test cases to collect function pairs whose lifetimes overlap, marking them as concurrent invocation candidates, and simultaneously records their respective neighboring functions as key points for delay injection. 

To achieve finer-grained thread scheduling and reduce randomness in thread interleaving,~SegFuzz \cite{jeong2023segfuzz} decomposes threads into instruction segments, 
mutates the instruction order within~each segment, and finally merges them into two concurrent threads that execute according to the mutated sequence. DDRace \cite{Yuan2023DDRace} targets more severe concurrency-related use-after-free (UAF) bugs~by~first identifying syscall sets that can trigger both free and use operations on the same memory region (rather than mere memory accesses), and then performing parameter-level mutation of a pseudo-syscall \texttt{setdelay} (which enables thread delay injection) based on newly designed feedback metrics (see Sec \ref{sec4.6}), such as determining which thread to delay and by what duration.

\subsubsection{A Combination of Generation and Mutation.}
In seed generation or mutation,~two~major~challenges often arise, \ie (i) the explosion of the seed space, and (ii) the tendency of random mutations~to disrupt the prerequisites, contexts, or complex system states required to reach deeper code regions. These issues not only undermine the performance of OSF (\eg new seeds struggle to explore deeper paths or trigger new coverage) but also affect its effectiveness in specific scenarios (\eg seeds being rejected early due to failing validation checks). Consequently, recent OSF research has increasingly focused on combining generation and mutation strategies to jointly address these challenges.

Hydra \cite{Hydra2020finding} first uses the semantic assistance of argument types to generate syscalls specific to file system operations (\eg \texttt{open} and \texttt{write}) and valid arguments (\eg correct file paths), which avoids rejection by error-checking code at an early stage. Building upon correct file system operations, Hydra employs argument-based mutators, based on AFL \cite{AFL}, to mutate the metadata blocks that are more likely to reveal file system vulnerabilities. HEALER \cite{sun2021healer} learns the dependency table~from the producer-consumer relationships in Syzlang, and utilizes it to guide the generation of syscall sequences with logical dependencies, thereby alleviating the syscall combination explosion problem caused by random generation. During the mutation phase, HEALER uses syscall-based mutation (\eg inserting syscalls) based on the dependency table, maintaining the dependencies within the sequence. PrIntFuzz \cite{ma2022printfuzz} extends syzlang by adding fault injection information (such as data, fault codes, and interrupt signals) to generate syscall sequences related to fault handling.~This~allows the mutation to be directly applied to useful test cases, avoiding prolonged trial and error caused by relying solely on mutations. To bypass the complex initialization checks in device drivers~(such as magic value verification and polling loops) and explore deeper code logic (such as packet~processing), Drifuzz \cite{shen2022drifuzz} combines concolic execution with forced execution to generate golden seeds that can successfully guide driver initialization. Starting from these golden seeds, Drifuzz, based on kAFL \cite{schumilo2017kafl}, applies argument mutators to the inputs to discover new execution paths. ViDeZZo~\cite{Liu2023ViDeZZoDV} generates context-dependent I/O messages based on the proposed lightweight grammars, and applies the proposed three types of mutators with different granularities to extend the diversity of messages, \ie intra-message mutators, inter-message mutators, and group-level mutators. These mutators are similar to the syscall, argument-based mutation, which essentially wraps and applies its basic mutators (\eg delete, change order, bit flip) to a specific domain. BRF \cite{Hung2024BRFFT} extracts eBPF domain knowledge (\eg validator rules) and syscall dependencies to generate semantically correct eBPF programs, and then mutates the syscalls on top of the generated programs, solving the problem that seeds generated by previous kernel fuzzers are difficult to pass the eBPF validator efficiently. To address the problem that random mutations often disrupt the constructed kernel state, MOCK \cite{Xu2024MOCKOK} employs a neural network model to dynamically learn the contextual dependencies within syscall sequences. Based on the trained model, MOCK predicts and generates new seeds~that~extend the current kernel state; for seeds that trigger new coverage, MOCK leverages the model~to~predict suitable insertion points and syscalls for mutation. SATURN \cite{Xu2024Saturn} first extracts the file operation structure of device drivers using kallsyms and kcov, achieving precise mapping of syscall sequences (\eg \texttt{ioctl\$printer}) and their parameters (\eg valid file paths like \texttt{/dev/device\_name}). This approach addresses the inefficiency of initial seeds in USB driver fuzzing. During the mutation phase, SATURN produces variants of these sequences through syscall-based and argument-based mutators, thereby exploring the input space for USB driver fuzzing.


\subsection{Feedback Mechanism}\label{sec4.6}

Feedback determines which seeds in the corpus are selected for mutation. We classify feedback~into code coverage, thread coverage, and custom feedback. Existing surveys \cite{Li2018FuzzingAS, Mans2018TheAS, Liang2018FuzzingSO} have systematically summarized code coverage in application fuzzing. Therefore, we briefly describe how code coverage is obtained in OSF and focus our discussion on thread coverage and custom feedback.

\subsubsection{Code Coverage.}
In OSF, code coverage is primarily obtained through kcov (see Sec \ref{subsection4.2}), which requires Linux 4.6 or later, gcc 6.1.0 or newer, or any Clang version supported by the kernel. It is worth mentioning that kcov cannot capture complete kernel coverage and has several limitations, such as the inability to monitor coverage during the driver initialization and enumeration phase when the kernel is not fully started, and the lack of support for collecting coverage from software and hardware interrupts. To address these problem, several studies \cite{schumilo2017kafl,wu2023devfuzz,ma2022printfuzz, zhao2022semantic,schumilo2021nyx} leverage Intel PT, which records branch instructions, calls, and target addresses, to compute code coverage and thereby capture instruction-level execution traces during program execution.

\subsubsection{Thread Coverage}

To effectively expose concurrency bugs, several studies \cite{xu2020krace,CONZZER2022context,jeong2023segfuzz,Yuan2023DDRace} have proposed various thread coverage metrics to explore multi-threaded scheduling interleaving.

Krace \cite{xu2020krace} proposes alias instruction pair (AIP) coverage, which is the first feedback mechanism related to thread coverage in OSF, as a metric of the degree of thread interleaving. An AIP~is~defined as a pair of consecutive instructions in two concurrent threads that read or write the same memory address. Formally, if there exists a $wi\_x$ instruction to write memory $A$ in thread $t\_1$, denoted as $A\leftarrow(wi\_x,t\_1)$, and there exists a consecutive $ri\_x$ instruction to read memory $A$ in another thread $t\_2$, denoted as $A\leftarrow(ri\_y,t\_2)$, then ($wi\_x \rightarrow ri\_y$) denotes an AIP. 
If a new AIP~is~discovered, the corresponding seed is considered valuable and subjected to thread scheduling perturbation.

However, AIP coverage ignores non-consecutive instructions and their dependent function call contexts. As a result, if no new AIP is triggered, potential data races may be missed \cite{CONZZER2022context}. 
To address this problem, CONZZER \cite{CONZZER2022context} introduces concurrency call pair (CCP) coverage, which records the complete call contexts of two concurrent functions to characterize thread interleaving. Specifically, a CCP consists of the call contexts of a pair of concurrent functions. 
By leveraging CCP, CONZZER can differentiate function call contexts to perturb thread scheduling, and thereby explore more potential data race scenarios, even for non-consecutive instructions.

Both AIP and CCP fail to distinguish fine-grained instruction-level interleaving when different read/write instructions access the same memory address (\eg $(wi\_3 \rightarrow ri\_1)$ and $(wi\_3 \rightarrow ri\_2)$). SegFuzz \cite{jeong2023segfuzz} addresses this limitation by introducing interleaving segment (IS) coverage, which decomposes a concrete thread interleaving into small instruction-level segments and uses them as coverage units. By capturing the relative execution order of multiple memory-accessing instructions on the same address, IS coverage enables finer-grained exploration of thread interleaving. 


For concurrent use-after-free (UAF) vulnerabilities, which involve not only concurrent accesses~to the same memory but also the special case of use after free, AIP lacks differentiation~of~instruction types and shared variables, CCP cannot distinguish specific shared variables accessed within a function, and IS fails to capture interleaving across multiple variables. As a result, they all struggle to capture the typical triggering scenario of concurrent UAF, where the free and subsequent use operations are often far apart and further depend on the state changes of other shared variables. To address this, DDRace \cite{Yuan2023DDRace} introduces race pair interleaving path (RPIP),~which~extends~AIP with a triplet <\textit{instruction type, thread ID, variable value}>, where \textit{instruction type} distinguishes~between read, write, and free operations, \textit{thread ID} identifies the executing thread to enable targeted scheduling perturbation in potential UAF regions, and \textit{variable value} captures variable state changes,~thereby~enabling the tracking of concurrency with cross-variable dependencies and condition-related triggers.


\subsubsection{Custom Feedback.}

To uncover specific scenarios or deeper vulnerabilities (\eg vulnerabilities that can only be triggered under specific kernel states even along already-covered paths \cite{zhao2022statefuzz}),~several works have designed target-specific coverage or feedback signal to guide the fuzzing.

For example, GREBE \cite{lin2022grebe} explores the potential behaviors of vulnerabilities in PoCs (\eg~a~general protection fault may evolve into a UAF) by applying taint analysis to identify critical kernel objects (\eg structure variables) and using them as coverage metrics.
Likewise, file system metadata~properties \cite{JANUS2019fuzzing}, variable state spaces \cite{zhao2022statefuzz}, and function error codes \cite{zhao2022semantic} are also quantified~as~coverage, guiding seed selection and mutation.
Besides, binary feedback signals are also utilized to provide a positive reward based on whether a seed triggers a specific target. 
For example, ACTOR \cite{fleischer2023actor}~and KOOBE \cite{chen2020koobe} summarize the triggering patterns of memory vulnerabilities, and introduce~action feedback (\eg UAF depends on the sequence \texttt{alloc} $\rightarrow$ \texttt{free} $\rightarrow$ \texttt{use}) and capability feedback (\eg the enhancement of out-of-bounds write capabilities, such as writing more bytes), respectively. VD-Guard \cite{Liu2023VDGuard} 
uses ``whether a DMA operation is triggered'' as a feedback signal to guide virtual device vulnerability exploration. DEVFUZZ \cite{wu2023devfuzz} leverages symbolic execution and static analysis to identify numerous validation constraints in Probe, MMIO, Port IO, and DMA paths, and uses~them as customized feedback; when seeds hit these constraints, mutations are applied to enable deeper driver vulnerability detection. Hydra \cite{Hydra2020finding} employs specific checkers (\eg SibylFS \cite{SibylFS2015} and B3 \cite{2018B3}), and treats detected file system bugs (such as POSIX violations and consistency bugs) as feedback signals to explore file system–related vulnerabilities.


\subsection{Bug Analysis}\label{Vulnerability}

OSF is designed to detect various types of bugs with the help of different monitors.

\subsubsection{Bug Types}

Bugs detected by OSF are categorized into four types, as shown in Table \ref{tab1},~including memory violation bugs (\textbf{M.}), concurrency bugs (\textbf{C.}), logic bugs (\textbf{L.}), and others (\textbf{O.}) that~include privilege protection, processor exception, or data integrity.

Memory violation bugs involve illegal access and manipulation of memory, including spatial~and temporal vulnerabilities. Spatial vulnerabilities refer to memory accesses that are outside the~scope~of their allocation, \eg OOB. Temporal vulnerabilities occur when memory~is~used at an incorrect moment, \eg UAF. Memory violation is the most common bug type in OSF, primarily because~C/C++ provides direct control over memory management but lacks security checking ~mechanism \cite{Unicorefuzz2019,chen2020koobe}. 

Concurrency bugs arise from insufficient synchronization in multi-threaded or multi-process environments, potentially causing data races, deadlocks, double-fetches, and other race conditions. Unlike user-space fuzzing, detecting such bugs in kernels is particularly challenging due to the large codebase, kernel complexity, and the non-deterministic nature of thread interleaving determined~by scheduling and synchronization primitives. These concurrency bugs may manifest as harmful races that lead to memory safety violations (\eg buffer overflows), which in turn can be exploited~\cite{jeong2019razzer}.

Logic bugs are caused by incorrect coding logic, e.g., virtual devices entering infinite loops when receiving invalid data \cite{henderson2017vdf,pan2021V-shuttle,Liu2023VDGuard}, functions being called multiple times when not designed~to~be reentrant (resulting in data inconsistencies) \cite{myung2022mundofuzz}, multiple releases~or~incorrect~formatting~in~interrupt request handling  \cite{wu2023devfuzz}, attempts to disable already disabled devices \cite{wu2023devfuzz}, divide-by-zero operations \cite{pan2021V-shuttle}, and assertion errors in the \texttt{BUG\_ON} macro \cite{wu2023devfuzz,2017DIFUZE,pailoor2018moonshine,schumilo2020hyper}.

In addition to the aforementioned three types, there are other relatively rare but equally critical OS bugs. For example, privilege protection vulnerabilities involve unauthorized operations~or~illegal access to protected resources, leading to sensitive data leakage or system integrity compromising; e.g., user space attempts to access high-privilege memory regions within the kernel \cite{wu2023devfuzz,2017DIFUZE}.~Processor exceptions occur when the processor attempts to execute undefined instructions or when the instruction encoding or structure does not conform to processor specifications, leading to abnormal system termination, e.g., invalid opcode \cite{wu2023devfuzz} and general protection faults~\cite{sun2021healer,lin2022grebe,maier2021bsod,zhao2022semantic}. Data integrity vulnerabilities occurs when data is inadvertently modified during storage or transmission, resulting in unusable or incorrect data states, e.g., data corruption \cite{pailoor2018moonshine}.

\subsubsection{Bug Monitor}

Muench et al. \cite{Muench2018Avatar2AM} track and categorize observable crashes and hangs into different classes, but this approach makes it difficult to detect vulnerabilities that do not immediately trigger a crash, such as buffer overflows. Another solution typically uses existing sanitizer tools \cite{Song2018SoKSF}  for monitoring, such as KASAN \cite{KernelAddressSanitizer2019}, UBSAN \cite{UBsan}, ASan \cite{ASAN}, and MSan \cite{MemorySanitizer2020}. 
When the~sanitizer~is not available (\eg in Hypervisor \cite{henderson2017vdf,schumilo2020hyper,schumilo2021nyx}), it is generally required to record a sequence of crash-causing test cases, and analysts debug to identify and categorize the bugs.

For concurrency bug monitoring, prior works \cite{Chen2020MUZZTG,Johansson2018RandomTW,Vinesh2019ConFuzzACF} integrate third-party checkers (\eg TSan \cite{TSan} and KCSAN \cite{KCSAN}) in fuzzing. However, such checkers have been reported with many false positives \cite{CONZZER2022context} due to the omission of special synchronization primitives such as message queues and conditional variables \cite{ThreadSanitizerDR}. To improve their precision, dynamic lockset analysis \cite{Savage1997EraserAD,jeong2019razzer,xu2020krace,jeong2023segfuzz}, happens-before analysis \cite{happensbefore}, and lockdep \cite{lockdep} are leveraged to customize the monitor.



\section{Distinctive Fuzzing Characteristics among OS Layers} \label{Section5}

Different OS layers, namely the kernel, file system, driver, and hypervisor, exhibit distinct fuzzing~requirements. These differences arise from the specific nature of their exposed testing interfaces~as~well as their respective roles within the OS architecture. We highlight the characteristic challenges at each layer, and review representative approaches that have been proposed to address them.

\subsection{Kernel Fuzzing}
Kernel fuzzers rely on syscall sequences as the testing interface, where variations in syscalls~and~their arguments may trigger different kernel behaviors. To conduct effective kernel fuzzing, it is essential to ensure the compliance of syscall sequences and argument usage, as well as their semantic~validity. Consequently, kernel fuzzing often attempts to address problems such as dependency inference, API polymorphism, and argument inference, as summarized in Table \ref{tab_kernel}.

\begin{table}
  \footnotesize 
  \caption{Kernel Fuzzers (Sorted by Publication Year)}
  \label{tab_kernel} 
  \vspace{-0.3cm}
  \begin{tabular}{ccccc}
  \toprule
  Fuzzers & OS & Dependency Inference & API Polymorphism & Argument Inference \\
  \midrule
  Moonshine\cite{pailoor2018moonshine} & Linux & TA \& CFA (ED, ID) & - & - \\
  HFL\cite{kim2020hfl} & Linux & PTA \& DSE (ED, ID) & DFA \& PTA \& DSE & CE (NS) \\
  HEALER\cite{sun2021healer} & Linux & SA \& DL (ED, ID) & - & - \\
  KSG\cite{sun2022ksg} & Linux & - & DH & SE (TC) \\
  FUZZNG\cite{bulekov2023FUZZNG} & Linux & - & - & HM (NS) \\
  SyzDirect\cite{tan2023syzdirect} & Linux & - & SA \& ICA & - \\
  KernelGPT\cite{yang2023kernelgpt} & Linux & - & - & LLM (TC) \\
  BRF\cite{Hung2024BRFFT} & Linux & DK (ID) & - & -\\
  MOCK\cite{Xu2024MOCKOK} & Linux & SA \& NN (ED, ID) & - & -\\
  \bottomrule 
  \end{tabular}
  \begin{flushleft}
  \justifying 
We use ``-'' to denote that it is irrelevant, or not mentioned, ``A (B)'' to denote that the fuzzer resolves~a~B-type issue using A technique, and ``A'' to denote that the fuzzer employs A technique to address the issue~of~API~polymorphism. Specifically, \textbf{TA}: Trace-based Analysis, \textbf{CFA}: Control Flow Analysis, \textbf{PTA}: Points-to Analysis, \textbf{DSE}: Dynamic~Symbolic Execution, \textbf{SA}: Static Analysis, \textbf{DL}: Dynamic Learning, \textbf{NN}: Neural Network, \textbf{DK}: Domain Knowledge, 
\textbf{DFA}: Data Flow Analysis, \textbf{DH}: Dynamic Hooking, \textbf{ICA}: Indirect Call Analysis, \textbf{CE}: Concolic Execution, \textbf{SE}: Symbolic Execution, \textbf{HM}: Hooking and Mapping, \textbf{LLM}: Large Language Model; \textbf{ED}: Explicit Dependency, \textbf{ID}: Implicit Dependency; \textbf{NS}: Nested Structure, \textbf{TC}: Type Casting. 
  \end{flushleft}
\end{table}



\subsubsection{Dependency Inference.}

Satisfying dependency relationships between syscalls is a key prerequisite for constructing valid syscall sequences (\ie seed behaviors). 
Such dependencies~are~generally divided into two categories. \textit{Explicit dependencies} occur when the result of a syscall $S_i$ is directly used as the input of another syscall $S_j$. 
\textit{Implicit dependencies} arise when $S_i$ influences~the execution~of $S_j$ through shared kernel data structures, even though no direct result transfer exists between them. 

Moonshine \cite{pailoor2018moonshine}, HFL \cite{kim2020hfl}, HEALER \cite{sun2021healer}, and MOCK \cite{Xu2024MOCKOK} all attempt to model both explicit and implicit dependencies. For explicit dependencies, HEALER and MOCK statically infer~syscall-level producer–consumer relationships during a preprocessing stage by leveraging the resource-based dependencies encoded in Syzlang \cite{syzlang}, where producer syscalls create resources that are subsequently consumed by consumer syscalls; e.g., \texttt{open} generates a file descriptor that is later consumed by \texttt{write}. To reduce the burden of manually maintaining Syzlang specifications, Moonshine abandons DSL-based inference of explicit dependencies, and instead leverages a customized Strace \cite{Strace} to trace pre-collected seed  and infer dependencies between syscall parameters and return values. However, the accuracy and completeness of the inferred dependencies are inherently constrained by the coverage and quality of the seed corpus. In contrast, HFL applies points-to analysis~at~the kernel function level to identify resource-based dependencies that point to the same kernel object;~e.g., the file descriptor returned by \texttt{socket} and the parameter of \texttt{connect} both point to the same struct socket object. However, this approach incurs substantial preprocessing overhead.

For implicit dependencies, HEALER infers them via dynamic learning by iteratively removing~adjacent syscalls and observing coverage changes. However, this method is limited to adjacent syscalls and cannot capture long-range dependencies. Moonshine applies static control flow analysis to examine read–write operations on shared data, thereby extending to cross-call dependencies.~However, as a purely static approach, it often produces false positives \cite{Hao2022DemystifyingTD}. 
HFL employs dynamic~symbolic execution to analyze path constraints and examine whether a syscall affects the execution path of subsequent syscalls by modifying kernel state, thereby improving the precision of dependency analysis. To mitigate state explosion, this approach restricts the number of explored execution~paths, which consequently imposes limitations on the scope of the analysis. MOCK introduces a neural language model (BiGRU) to learn contextual information among syscalls, thereby enhancing its capability to capture implicit dependencies across syscalls. The effectiveness of this approach~depends on the quality of the initial seed corpus and the intermediate seeds generated during fuzzing.

In domain-specific fuzzing, syscall sequences need to be constrained according to the specialized execution workflows and testing requirements of the target domain. In the~eBPF~domain, additional syscalls are generated to attach eBPF programs to appropriate kernel hooks~and~to trigger their execution (\eg invoking \texttt{BPF\_PROG\_ATTACH} and \texttt{BPF\_PROG\_TEST\_RUN} within syscall sequences).

\subsubsection{API Polymorphism.}
In the syscall execution, polymorphism refers to the ability of a single syscall interface to invoke different kernel handler functions depending on its input arguments. It provides a unified entry point, avoiding the need to introduce separate syscalls for each functionality. For example, \texttt{ioctl} triggers different handler functions based on the \texttt{cmd} argument. 
Considering nearly 4,200 syscall variants in Linux \cite{Syzkaller}, ignoring polymorphism can easily lead to the problem where test inputs fail to reach the expected code paths \cite{jeong2019razzer,schumilo2017kafl}.

To address this problem, HFL \cite{kim2020hfl} uses inter-procedural data flow analysis and static points-to analysis to determine which index variables of function~pointer tables originate from syscall~arguments. Based on this, HFL implements an offline converter that expands the calls of function~pointer tables into explicit conditional branches. Finally, HFL employs symbolic execution on the converted explicit conditional branch paths to explore the syscalls and argument values that trigger the target functions. However, function pointer tables may be dynamically registered during kernel module initialization, dynamic configuration by other syscalls, and device hot-swapping (\eg different~protocols like TCP can register specific operations at various times), making static analysis unable to capture all control flow paths completely. To address the problem of dynamic registration, KSG dynamically hooks multiple probes before and after specific kernel functions. By scanning device files and network protocols, KSG triggers the execution of hooked kernel functions to extract the syscall entry points of triggered submodules. This method introduces additional runtime overhead, especially when handling a large number of syscalls and frequent dynamic registrations. Moreover, since KSG relies on runtime behavior, it may miss some unregistered function pointers, resulting in incomplete analysis. SyzDirect uses static analysis to identify and locate anchor functions to reduce the substantial overhead of modeling all functions, and combines type-based indirect call analysis to trace indirect call chains to determine syscall variants and arguments. 


\subsubsection{Argument Inference.}

Invalid argument can lead to ineffective seeds that fail to explore the intended execution paths. Hence, argument inference is essential in kernel fuzzing, and is typically divided into argument type inference and nested structure inference.

Argument type casting is common in kernel code. For the sake of generality, many syscalls~declare their parameters as abstract pointer types. 
However, during execution, these arguments are often interpreted as different data structures depending on runtime conditions. 
To address this problem, KSG utilizes symbolic execution to perform path-sensitive analysis on the target function to check for the presence of type casting operations on the arguments (scalar-to-pointer and pointer-to-pointer casting). A global mapping table is constructed for symbols and memory regions, which solves the problem that type casting is difficult to inference due to alias propagation. 
KernelGPT prompts GPT-4 to automatically infer the unknown types of arguments and the specific values of arguments. However, it is currently limited to the argument type inference of \texttt{ioctl} syscalls only.

A nested structure refers to a structure whose field members point to another structure that is dynamically allocated and initialized at runtime. 
Such nested structures are usually frequent in special kernel APIs like \texttt{copy\_from\_user} and \texttt{copy\_to\_user} that are used for memory address~copying, and traditional fuzzing have a hard time inferring such complex argument structure due to dynamic allocation of addresses in such cases. Similarly, symbolic execution is based only on static analysis and explicitly symbolized input space; it cannot handle dynamic memory regions~pointed to by nested pointers. In order to accurately trace the memory operations of nested structure pointers, HFL specifically instruments the \texttt{copy\_from\_user} and \texttt{copy\_to\_user} functions, and captures~their arguments and return values (\ie the addresses and sizes of the source and target buffers). Additionally, the intercepted arguments are labeled as symbolic variables, and the buffer addresses of the nested structures as well as the data lengths are inferred using concolic execution.~It~traces~the propagation paths of arguments to constrain syscall and its arguments mutations, but relies specifically on handwritten Syzlang, and covers limited syscalls and requires a lot of manual work.
To address this problem, FUZZNG performs checksum corrections on mutated syscalls to ensure the validity of the syscall sequence. Specifically, FUZZNG develops a kernel module mod-NG that hooks the kernel's APIs related to handling file descriptor allocation and pointer arguments to reshape the input space of syscalls. For file descriptors, mod-NG hooks the kernel's \texttt{alloc\_fd} API, which is used to allocate new file descriptors, and the \texttt{fdget} API, which is used to retrieve file objects via file descriptors. By intercepting these APIs, mod-NG maps invalid file descriptors generated by mutations to existing file objects within the kernel. For pointer arguments, mod-NG hooks the \texttt{copy\_from\_user} function, allowing the kernel to populate the structure pointed to by the pointer into a valid user-space memory region when accessing user-space memory. Through this approach, FUZZNG reshapes the input space of syscalls, ensuring that even if the mutated syscall arguments are invalid, they can still be mapped to valid file objects and memory regions.

\subsection{File System Fuzzing}


Beyond ensuring the correctness of individual file operations, a file system must also maintain~a~correct persistent status across time, across processes, and in the presence of system crashes. This~status correctness is fundamentally manifested in the consistency among file system metadata as well~as between metadata and the persistent data status they describe (\eg the consistency between inodes, allocation bitmaps, and the actual allocation of data blocks). Consequently, effective file system fuzzing must jointly exercise file operation sequences and file system status evolution to expose vulnerabilities that stem from persistent status. Therefore, we review existing file system fuzzers along three dimensions, \ie input type, status, and specific bugs, as summarized in Table~\ref{tab_filesystem}.

\subsubsection{Input Type}
File system fuzzers can be categorized into syscall- and syscall+image-based according to the input types. File systems are mounted to the kernel as files. Therefore,~file system fuzzers can fuzz file system-related source code in the kernel by generating only syscall sequences that specialize in operating on files (\eg Syzkaller \cite{Syzkaller}, kAFL \cite{schumilo2017kafl}, Krace \cite{xu2020krace}, CONZZER \cite{CONZZER2022context}, and Monarch\cite{TAO2024MONARCH}). 
However, syscall interfaces expose only high-level semantic parameters (\eg file paths and flags), while persistent metadata allocation and updates are implicitly determined~by internal file system logic. As a result, syscall-only inputs do not provide explicit or controllable~access to the metadata that governs the evolving runtime status of a mounted file system, which~typically occupies only a small fraction of the entire file system image (\eg around 1\% \cite{JANUS2019fuzzing}). Consequently, syscall-only fuzzers often fail to preserve semantic dependencies and object lifetimes across operations, leading to semantically invalid executions that do not exercise meaningful file system behaviors (\eg issuing \texttt{read()} or \texttt{unlink()} operations on files already modified by prior \texttt{rename()} operation). To address this limitation, JANUS \cite{JANUS2019fuzzing}, Hydra \cite{Hydra2020finding}, and Lfuzz \cite{Liu2023LFuzz} adopt both the file system image (enabling direct, fine-grained control over persistent metadata) and the syscalls operating on it as inputs, thereby enabling persistent-status–aware file system fuzzing.


\subsubsection{Status} The file system status refers to an abstract representation of the runtime status of file objects on a mounted image, including their existence, types, namespace relationships, and accessibility, which collectively determine the set of semantically valid file operations. 
Syscall-only fuzzers do not consider the set of file operations that are semantically permitted by the current file system status induced by the underlying image (metadata). Instead, they rely on coarse-grained control via the syscall interfaces and blindly mutate file objects (\eg directories and symbolic links) \cite{JANUS2019fuzzing}, which may arbitrarily perturb the file system status and lead to testing behaviors that execute file operations in semantically invalid ways. In contrast, syscall+image fuzzers maintain intermediate status on a controllable image throughout the fuzzing process and jointly mutate the image status and the associated file operations, thereby generating status-aware test executions.

JANUS \cite{JANUS2019fuzzing} is the first file system fuzzer to correlate status with file operations. 
It first parses a controlled file system image to extract metadata blocks, and inspects the image to enumerate file objects and derive their initial runtime status. File operations (\ie syscall sequences) are generated and mutated under the constraints imposed by the maintained status, which is incrementally~updated as metadata and operations evolve. During mutation, JANUS perturbs the metadata blob (\eg affecting object types or inode sizes) and coordinately mutates file operations within the space~permitted by the updated status (\eg avoiding calls that conflict with previously removed~objects).~Lfuzz~\cite{Liu2023LFuzz} extends JANUS by tracking recently accessed metadata locations and their spatial~neighbors, focusing metadata mutations on regions more likely to be exercised by subsequent file operations, 
thereby alleviating inefficiencies incurred by coordinated fuzzing over a large metadata mutation space.


Hydra~\cite{Hydra2020finding} reveals that the dependency of file operation behavior on metadata status may~undergo a shift in dominance. Specifically, even if a fuzzer records and maintains a semantically~correct~status, the constraining power of this status over subsequent file operations may gradually give way to kernel runtime status accumulated through repeated file operations. 
As a result,~bug reproduction becomes increasingly dependent on kernel runtime statues that are difficult to precisely control and reproduce, thereby weakening the causal linkage between metadata mutations and bug manifestation. To mitigate the accumulation of kernel runtime status and improve bug reproducibility,~Hydra employs a library OS–based executor (\eg the Linux Kernel Library) to create a fresh execution~instance for each test inpus. 
In addition, Hydra adopts a staged mutation scheduler that prioritizes metadata-level image mutations over immediately extending syscall sequences, which helps, to some extent, maintain controllability over the evolution of file system status.


\begin{table}
    \centering
    \footnotesize 
    \caption{File System Fuzzers (Sorted by Publication Year)}
    \label{tab_filesystem}
    \vspace{-0.3cm}
    \begin{tabular}{ccccc}
    \toprule
    \multicolumn{1}{m{2cm}}{\centering Fuzzers} &  
    \multicolumn{1}{m{4cm}}{\centering File System} &
    \multicolumn{1}{m{3cm}}{\centering Input Type} &
    \multicolumn{1}{m{0.5cm}}{\centering Status} &
    \multicolumn{1}{m{1.8cm}}{\centering Specific Bugs} \\
    \midrule
    Syzkaller\cite{Syzkaller} & ext4, Btrfs, F2FS, jfs, xfs, reiserFS & syscall & $\times$ & C\\
    kAFL\cite{schumilo2017kafl} & ext4 & syscall & $\times$ & -\\
    JANUS\cite{JANUS2019fuzzing} & ext4, Btrfs, F2FS& syscall+image & \checkmark & L\\
    Krace\cite{xu2020krace} & ext4, Btrfs & syscall & $\times$ & C\\
    Hydra\cite{Hydra2020finding} & ext4, Btrfs, F2FS & syscall+image & \checkmark & I,S,C,L\\
    CONZZER\cite{CONZZER2022context} & Btrfs, jfs, xfs, reiserFS & syscall & $\times$ & C\\
    Lfuzz\cite{Liu2023LFuzz} & ext4, Btrfs, F2FS & syscall+image & \checkmark & L\\
    Monarch\cite{TAO2024MONARCH} & Lustre, GlusterFS, NFS & syscall & $\times$ & I,C,L\\ 
    \bottomrule
    \end{tabular}
    \begin{flushleft}
      \justifying 
      We use ``-'' to denote that it is irrelevant, or not mentioned. Specifically, \textbf{I}: Crash Inconsistency, \textbf{S}: Specification Violation, \textbf{C}: Concurrency Bug, and \textbf{L}: Logic Bug.
      \end{flushleft}
\end{table}

\subsubsection{Specific Bugs}

Compared to other OS layers, bugs in the file system are characterized~by~diversity and specificity \cite{Lu2013ASO}. In addition to mainstream memory violation errors, bugs specific~to~file systems include crash inconsistency, specification violation, concurrency bug, and logic bug due to their intrinsic data persistence and concurrency characteristics. 

\textit{Crash inconsistency} is the most typical vulnerability in file system, which refers to a file system fails to recover to a correct status after a system crash caused by a power loss or a kernel panic. This is typically due to the file system not correctly managing its metadata when writing, updating, or deleting data, leading to the catastrophic consequences of permanent loss or corruption~of~files. \textit{Specification violation} occurs when a file system deviates from the standards or specifications~it is expected to follow during operation. File system interfaces are typically governed by well-defined specifications, such as the POSIX standard or Linux man pages, which prescribe the~allowed~behaviors and error codes of file operations.  For example, the POSIX standard specifies that when executing the \texttt{unlink} syscall, the only permissible error code is \texttt{EPERM}~\cite{Hydra2020finding}; however, some file system implementations return \texttt{EISDIR}, thereby violating the POSIX specification. Such bugs can severely undermine the robustness and security of software built upon file systems (\eg due to improper error handling). Moreover, because they typically do not lead to crashes, they are difficult to capture in practice. Consequently, the capability to detect such bugs remains limited beyond regression testing and SibylFS~\cite{SibylFS2015}, on which Hydra’s specification-violation~checker~\cite{Hydra2020finding}~is~built. 
Although \textit{concurrency bug} is not specific to file systems, modern file systems introduce a range of programming paradigms to exploit multi-core computing~\cite{fstests} (\eg read-copy-update (RCU) and asynchronous work queues), which improve performance but significantly increase the likelihood of writing error-prone concurrent code.
Unlike other types of bugs, file system-specific \textit{logic bugs} do not adhere to any general pattern for definition (\eg F2FS requires its own concept of rb-tree consistency \cite{Hydra2020finding}). As discussed in Section \ref{Vulnerability}, logic bugs do not cause immediate crashes but can lead to undefined behavior over time, affecting both performance and reliability.

\subsection{Driver Fuzzing}
Drivers interact directly with hardware and constitute the largest and structurally most complex subsystem in the kernel, which makes driver fuzzing exhibit several design dimensions that are distinct from kernel and file system fuzzing. Table \ref{tab_driver} categorizes existing driver fuzzers along~these dimensions, \ie input type (additional hardware-side attack surfaces), the covered~stages (driver processing workflow), and whether device-free execution is supported (scalability consideration).

\subsubsection{Input Type}

As kernel subsystems, driver can also be tested by fuzzing syscalls. For example, in Linux, each file under the /dev directory represents a hardware device. User-space~applications can obtain a file descriptor for a device and interact with it using syscalls such as \texttt{read}, \texttt{write}, or \texttt{ioctl} to perform specific hardware operations. Existing fuzzers, such as usb-fuzzer \cite{Syzkaller} (the driver fuzzing module in Syzkaller), DIFUZE \cite{2017DIFUZE}, BSOD \cite{maier2021bsod}, StateFuzz \cite{zhao2022statefuzz}, SyzDescribe~\cite{hao2023syzdescribe}, SATURN \cite{Xu2024Saturn}, and Syzgen++ \cite{Chen2024SyzGen++}, are all based on Syzkaller to generate syscall sequences that~interact with drivers.
To enable driver fuzzing on real, mass-produced Android devices running signed kernels, a driver fuzzer must avoid breaking the device’s original secure boot chain. Consequently, Ex-vivo \cite{Pustogarov2020ExvivoDA} adopts AFL to generate test inputs for syscalls without requiring kernel modifications, rather than using Syzkaller, which relies on a customized, unsigned kernel for instrumentation.


However, it is insufficient to consider syscalls as the sole entry point for fuzzing, which 
overlooks the complexity and potential bugs in hardware-side code, resulting in inadequate coverage~and attention to hardware-level bugs \cite{peng2020usbfuzz}. Therefore, another fuzzing interface involves injecting~malicious inputs from the hardware side through device configuration or I/O such as Port I/O, MMIO, and DMA. USBFuzz \cite{peng2020usbfuzz} uses a simulated USB device to match with kernel drivers, and injects malicious USB descriptors. Fuzzers such as \cite{Song2019PeriScopeAE,song2020agamotto,zhao2022semantic,ma2022printfuzz,shen2022drifuzz,wu2023devfuzz,Jang2023ReUSB, Huster2024ToBoldly} intercept the I/O~access of target devices (including the target address and data content). After mutating the I/O data, these fuzzers resend the mutated data to the corresponding I/O. This approach enhances the flexibility of driver fuzzers, and improves their ability to discover device-side bugs, such as those associated with hardware initialization, interrupt handling, direct DMA operations, and hardware error handling. 

\begin{table} 
    \centering 
    \scriptsize 
    \caption{Driver Fuzzers (Sorted by Publication Year)} 
    \label{tab_driver} 
    \vspace{-0.3cm}
    
    \begin{tabular}{ccccc} 
    
    \toprule 
    \multicolumn{1}{m{2cm}}{\centering Fuzzers} &  
    \multicolumn{1}{m{3.5cm}}{\centering Device Type} &
    \multicolumn{1}{m{2cm}}{\centering Input Type} &
    \multicolumn{1}{m{2cm}}{\centering Stage} &
    \multicolumn{1}{m{2cm}}{\centering Device Dependence}\\
    
    \midrule 
    \multirow{1}*{usb-fuzzer\cite{Syzkaller}} & USB & syscall & 3 & physical device  \\

    \multirow{1}*{DIFUZE\cite{2017DIFUZE}} & PCI/USB/I2C/Others & syscall & 3 & physical device  \\

    \multirow{1}*{Periscope\cite{Song2019PeriScopeAE}} & PCI/USB/I2C/Others & I/O & 3 & physical device  \\

    \multirow{1}*{USBFuzz\cite{peng2020usbfuzz}} & USB & device configuration & 1, 2, 3 & device behavior \\

    \multirow{1}*{Agamotto\cite{song2020agamotto}} & PCI/USB & syscall+I/O & 3 & physical device \\

    \multirow{1}*{Ex-vivo\cite{Pustogarov2020ExvivoDA}} & PCI/USB/I2C/Others & syscall & 1, 2, 3 & device free \\

    \multirow{1}*{BSOD\cite{maier2021bsod}} & PCI & syscall & 1, 2, 3 & device behavior \\

    \multirow{1}*{StateFuzz\cite{zhao2022statefuzz}} & Others & syscall & 3 & physical device \\

    \multirow{1}*{Dr.Fuzz\cite{zhao2022semantic}} & PCI/USB/I2C/Others & I/O & 1, 2, 3 & device free \\

    \multirow{1}*{PrintFuzz\cite{ma2022printfuzz}} & PCI/USB/I2C/Others & syscall+I/O & 1,2,3 & device behavior \\

    \multirow{1}*{DriFuzz\cite{shen2022drifuzz}} & PCI/USB & I/O & 1, 2, 3 & device free \\

    \multirow{1}*{SyzDescribe\cite{hao2023syzdescribe}} & PCI/USB/I2C/Others & syscall & 3 & physical device \\

    \multirow{1}*{DEVFUZZ\cite{wu2023devfuzz}} & PCI/USB/I2C/Others & I/O & 1, 2, 3 & device free \\

    \multirow{1}*{ReUSB\cite{Jang2023ReUSB}} & USB & syscall+I/O & 1, 2, 3 & device behavior \\
    
    \multirow{1}*{SATURN\cite{Xu2024Saturn}} & USB & syscall & 1, 2, 3 & device behavior \\

    \multirow{1}*{Syzgen++\cite{Chen2024SyzGen++}} & Others & syscall & 3 & physical device \\

    \multirow{1}*{VIRTFUZZ\cite{Huster2024ToBoldly}} & PCI & I/O & 1, 2, 3 & device behavior \\
    
    \bottomrule
    \end{tabular}
    \end{table}

\subsubsection{Stage}
From initialization to the ready state, a device driver typically progresses through~three conceptual stages, \ie device enumeration, probe execution, and device communication. The~first~two stages collectively form the validation chain, which enforces a series of input-dependent checks, such as validating the chip revision, determining the I/O mechanism, and verifying vendor and product identifiers. 
In practice, fuzzing approaches pass the validation chain either by assuming well-formed device behavior—via real hardware (using physical devices \cite{Song2019PeriScopeAE} or already device-bound drivers \cite{Syzkaller,2017DIFUZE,song2020agamotto,zhao2022statefuzz,Chen2024SyzGen++,hao2023syzdescribe}) or high-fidelity device behavior (\eg emulation \cite{peng2020usbfuzz,ma2022printfuzz,Xu2024Saturn} and record-and-replay \cite{Jang2023ReUSB,maier2021bsod,Huster2024ToBoldly})—or by tools specifically designed to construct inputs that satisfy validation requirements without relying on predefined device behavior (\eg interception–forgery \cite{Pustogarov2020ExvivoDA}, concolic execution \cite{shen2022drifuzz} or feedback-aware \cite{zhao2022semantic}).  
These approaches are primarily designed to ensure successful traversal of the validation chain, and consequently focus on communication-related code regions. Another approach, DEVFUZZ \cite{wu2023devfuzz}, considers how to correctly pass the~validation chain while additionally fuzzing the three stages separately and adequately.

\textbf{The Validation Chain (\ie Stage 1 and 2)} determines whether a device can be matched~with and successfully claimed by a certain class of drivers in the kernel, and thus constitutes a prerequisite for effective driver fuzzing. Hardware-based fuzzers rely on real physical devices, whose device enumeration and probe execution naturally pass the validation chain checks at the physical level.

High-fidelity device emulation approaches approximate the behavior of real hardware at the~software level, enabling the kernel to perceive the presence of a device and thereby undergo the~same enumeration and driver-matching process. These approaches are typically built upon QEMU \cite{bellard2005qemu}, which provides multi-bus device models and a unified device abstraction to construct virtual~devices, and leverage runtime hot-plug mechanisms to trigger device enumeration and probe execution in the kernel \cite{peng2020usbfuzz, ma2022printfuzz}.
In addition, the Linux kernel provides the USB gadget framework, which implements device-side logic directly within the kernel. Unlike QEMU-based approaches, the gadget framework does not emulate hardware devices explicitly. Instead, it realizes firmware behavior that would normally run on external devices as in-kernel gadget functions, and triggers device enumeration at runtime via command-line interfaces, thereby invoking the probe logic of the corresponding drivers \cite{Xu2024Saturn}. Compared to QEMU, the gadget framework substantially reduces the complexity of device-side modeling and allows fuzzing to focus on higher-level USB driver interaction logic. This simplification, however, comes at the cost of reduced coverage of diverse device configurations and low-level hardware-dependent execution paths.
Fuzzers based on high-fidelity~record-and-replay \cite{Jang2023ReUSB,maier2021bsod,Huster2024ToBoldly} take a different approach; \ie they capture interaction traces between real devices~and drivers in a physical environment and subsequently replay these interactions in a device-free~setting, thereby satisfying the validation chain requirements without requiring live hardware.

Besides, fuzzers specifically design new algorithms or software mechanisms to explicitly pass the validation chain. Ex-vivo \cite{Pustogarov2020ExvivoDA} intercepts validation checks and forges values that satisfy these checks, allowing syscall interfaces such as ioctl to bypass the validation chain even in the absence of real hardware. DriFuzz \cite{shen2022drifuzz} relies only on QEMU’s hot-plug mechanism to trigger~device~enumeration; unlike \cite{peng2020usbfuzz, ma2022printfuzz}, it does not construct complete device behavior for validation. Instead, it combines concolic execution with forced execution to identify values that satisfy magic checks, polling loops, and version checks within the validation chain. Dr.Fuzz \cite{zhao2022semantic} fixes bus scanning via snapshots as the fuzzing entry point, and leverages error codes as a new feedback signal to guide progress through the validation chain. DEVFUZZ \cite{wu2023devfuzz} treats the validation chain itself as the fuzzing target, which combines symbolic execution to generate device information that can pass probe checks and repeatedly triggers device enumeration by re-executing bus rescan–related \texttt{echo} commands. Notably, DEVFUZZ employs Intel PT \cite{kleen2015} to compensate for the lack of kcov \cite{vyukov2018kcov} coverage during the probe phase, where kcov is not yet available.

\textbf{Device Communication (\ie Stage 3)} 
continuously handles requests from the user-side~(primarily via syscall) as well as device-side I/O (MMIO/PIO/DMA/IRQ). The driver mainly executes~control-related and data- and asynchronous-related code logic. Control-related~functionality~is~responsible for configuring, managing, and transitioning device and driver states, without directly handling~bulk data transfer. Data- and asynchronous-related functionality processes device-side inputs and asynchronous events, including data transfer, interrupt handling, and deferred execution.

Typically, syscall-based fuzzers focus on generating valid syscall arguments to exercise control-related functionality. 
In contrast, I/O-based fuzzers focus on generating bus-compliant I/O messages to cover hardware-proximal data and asynchronous logic. 
Moreover, some driver bugs arise from the interleaved execution of control logic and data/asynchronous processing paths, and relying on~a single input dimension is often insufficient to expose such bugs. To this end, several works~jointly~use syscalls to drive the driver into specific runtime states while simultaneously mutating device-side I/O inputs, thereby covering bugs that span different input dimensions, \eg UAF or race conditions caused by the concurrency between incorrect state configurations and asynchronous I/O events.

\subsubsection{Device Dependence}

Existing fuzzers can be broadly categorized into three classes according to the degree of dependence on physical devices, \ie physical device, device behavior, and device-free.
\textit{Physical device–based fuzzers} (also referred to as hardware-in-the-loop testing) conduct fuzzing in real hardware environments \cite{Song2019PeriScopeAE} or against drivers that are already bound to physical~devices~\cite{Syzkaller,2017DIFUZE,song2020agamotto,zhao2022statefuzz,Chen2024SyzGen++,hao2023syzdescribe}. Such approaches offer high fidelity, naturally reflecting real device behavior without requiring explicit modeling of device protocols. However, their limited flexibility makes it difficult to scale them to broadly test large driver codebases. 
\textit{Device behavior–based fuzzers}~\cite{peng2020usbfuzz,ma2022printfuzz,Xu2024Saturn,Jang2023ReUSB,maier2021bsod,Huster2024ToBoldly} do not require real hardware; however, their effectiveness is inherently constrained by the need for device-specific expertise or by the substantial overhead of modeling and recording device behavior. As a result, such approaches face limitations in terms of flexibility and scalability, and extending them to drivers across different bus types or device categories typically requires significant additional effort.
\textit{Device-free approaches} \cite{Pustogarov2020ExvivoDA,shen2022drifuzz,zhao2022semantic,wu2023devfuzz} enable driver fuzzing without requiring real devices or complete device behavior modeling by designing specialized tools and algorithms. For example, some works perform explicit analysis or instrumentation of driver source code to enable source-level symbolic execution \cite{wu2023devfuzz}, or introduce novel feedback mechanisms \cite{shen2022drifuzz}, thereby achieving device free driver fuzzing in a source-aware manner.



\subsection{Hypervisor Fuzzing}

To fuzz a hypervisor, it is necessary to generate privileged operations that can trigger VM-exit and subsequently exercise internal hypervisor code paths. Accordingly, we summarize existing fuzzers along two dimensions in Table \ref{tab_hypervisor}, namely the types of interfaces that trigger VM-exit (\ie input~type) and the methodologies used to construct hypervisor fuzzing frameworks (\ie methodology).

    
    
    







    

\begin{table}
    \centering
    \scriptsize 
    \caption{Hypervisor Fuzzers (Sorted by Publication Year)}
    \label{tab_hypervisor}
    \vspace{-0.3cm}
    \begin{tabular}{cccc}
    \toprule
    Fuzzers & Hypervisor & Input Type & Methodology \\
    \midrule
    VDF\cite{henderson2017vdf} & QEMU & PIO, MMIO & Behavior mediation \\
    Hyper-Cube\cite{schumilo2020hyper} & \makecell{QEMU, Bhyve, ACRN, \\ VirtualBox, Vmware Fusion} & PIO, MMIO, DMA, Hypercall, Instruction & Bytecode translation \\
    NYX\cite{schumilo2021nyx} & QEMU, Bhyve & PIO, MMIO, DMA, Hypercall, Instruction & Bytecode translation \\
    V-Shuttle\cite{pan2021V-shuttle} & QEMU, Virtual Box & DMA & Hijacking \\
    MundoFuzz\cite{myung2022mundofuzz} & QEMU, Bhyve & PIO, MMIO, DMA & Hijacking \\
    Morphuzz\cite{bulekov2022morphuzz} & QEMU, Bhyve & PIO, MMIO, DMA & Hijacking \\
    IRIS\cite{cesarano2023iris} & Xen & Instruction & Hijacking \\
    VD-Guard\cite{Liu2023VDGuard} & QEMU, VirtualBox & PIO, MMIO, DMA & Behavior mediation \\
    ViDeZZo\cite{Liu2023ViDeZZoDV} & QEMU, VirtualBox & PIO, MMIO, DMA & Hijacking\\
    HYPERPILL\cite{Bulekov2024HYPERPILLFF} & QEMU & PIO, MMIO, DMA, Hypercall & Hijacking\\
    \bottomrule
    \end{tabular}
\end{table}

\subsubsection{Input Type}

Existing fuzzers leverage three categories of interfaces to trigger VM-exit, \ie I/O, hypercall, and privileged instruction. 
\textit{I/O} serves as the interface connecting the guest and the device emulation code within the hypervisor. For example, in an emulated-device environment~(\eg a virtual network interface card), the guest driver prepares packet buffers and descriptors (\eg destination address and size) in guest memory, and writes the physical address of this memory region to device registers via PIO or MMIO, thereby triggering a VM-exit. The hypervisor then~enters the device emulation routine, where it interprets register values, validates requests, and updates~the device state machine, followed by accessing guest memory through software-emulated DMA  to complete packet transmission. Upon completion, a virtual interrupt is injected into the guest~to notify the event. Through this workflow, the virtual machine monitor fully reproduces the operational semantics of physical devices in software, ensuring the correctness of device behavior~within the guest. \textit{Hypercall} provides virtual resource management interfaces that are independent~of~device emulation, 
and allows the guest to actively trigger VM-exit and explicitly request the hypervisor to perform privileged resource management operations. For example, the guest may use hypercall to provide the hypervisor with scheduling hints indicating that a virtual CPU is idle, thereby assisting the hypervisor in reallocating processor time. Moreover, architecture-specific \textit{privileged~instruction} 
cannot directly modify the physical hardware state within a hardware-isolated virtualized environment. Instead, the execution of such instructions by the guest triggers a VM-exit. The hypervisor intercepts this event and updates the corresponding virtual CPU state 
based on the instruction's semantics, ensuring the guest observes execution behavior consistent with bare-metal hardware. Since these privileged instructions directly drive the hypervisor's control logic concerning address space and execution context management, they constitute a critical interface for fuzzing.


\subsubsection{Methodology}

\textit{Behavior mediation-based approaches} construct or reconstruct realistic I/O~sequences using dedicated techniques, treating them as tester-defined behavior descriptions, and leverage specialized testing frameworks to materialize these descriptions into executable test cases that trigger real I/O instructions. Such testing frameworks include both hypervisor-internal mechanisms (\eg QTest \cite{QTest} in QEMU) and tester-implemented execution carriers (\ie harnesses).~These frameworks simulate vCPU behavior and directly issue PIO/MMIO read and write requests to virtual devices. 
As a result, when targeting non-QEMU hypervisors, testers typically need to implement equivalent harnesses to support the replay and execution of I/O behaviors. Under this~paradigm, fuzzers primarily focus on generating semantically valid I/O behaviors, including correct I/O dependencies, memory addresses, read/write instruction types, and data payloads that conform to device protocols. For example, VDF \cite{henderson2017vdf} adopts a record-and-replay mechanism to mutate and~replay test inputs derived from real I/O behavior traces. In contrast, VD-Guard \cite{Liu2023VDGuard} leverages static analysis to identify behavioral dependencies within virtual devices, from PIO/MMIO entry points to DMA scheduling sites, thereby constructing I/O behavior chains that can trigger DMA operations.

\textit{Bytecode translation-based approaches} are required to introduce a customized guest OS, which is specifically responsible for translating the random byte streams generated by the fuzzer into concrete operations that can trigger VM-exit, including I/O, hypercall, and privileged instruction, thereby enabling cross-platform compatibility for hypervisor fuzzing. Hyper-Cube \cite{schumilo2020hyper} randomly generates bytecode; however, due to its lack of awareness of internal virtual device state machine dependencies (\eg attempting to read a resource before it has been opened), it often generates a large number of semantically invalid I/O sequences. NYX \cite{schumilo2021nyx} introduces affine types to restrict each variable to be used at most once. By preventing the use of uninitialized resources or incorrect handle references, it addresses the strict ordering and data dependency requirements in I/O sequences.

\textit{Hijacking-based approaches} circumvent the difficulty of constructing valid inputs or complete execution paths from scratch by hijacking the interaction boundary between the hypervisor and the guest. Such approaches intercept I/O-related APIs via hooks \cite{myung2022mundofuzz,pan2021V-shuttle,bulekov2022morphuzz,Liu2023ViDeZZoDV}, and inject mutated values into device registers (PIO/MMIO), guest memory (DMA), as well as calling-convention~registers used to pass hypercall identifiers and parameters to drive testing. Alternatively, they directly hijack the hypervisor’s VM-exit handler \cite{cesarano2023iris} and start fuzzing from the VM-exit handling stage, or repeatedly restore the hypervisor to a state that is ``about to handle a VM-exit'' using snapshots \cite{Bulekov2024HYPERPILLFF}, thereby bypassing traditional testing approaches that rely on guest-internal execution paths and directly driving the hypervisor’s trap-and-emulate logic.



\section{Future Research Directions}\label{Section7}

\textbf{LLM-Based Test Case Generation.} Despite the widespread recognition of Syzkaller in kernel fuzzing, its support for rapidly iterating and continuously integrating open-source OSs exhibits~significant limitations. Particularly,~for~systems such as Linux, with an average of 200 commits~to~the mainline per day \cite{Hao2022DemystifyingTD}, Syzlang necessitates continuous manual updates and enhancements. This is not only inefficient, but also leads to limited code coverage due to incomplete syscall descriptions. The advent of large language models (LLMs) for automatic test case generation proposes a potential solution, leveraging the formidable language processing capabilities of LLMs for adapting OS specifications to Syzlang language learning and fine-tuning. This approach not only paves~new avenues for the automatic generation of Syzlang, but also aims to enhance the accuracy and coverage of test cases. Moreover, the linguistic processing prowess of LLMs could be employed to precisely analyze complex driver modules, such as DMA communications, thereby facilitating~the~automatic generation of test cases for Driver and Hypervisor layer I/O interfaces. Research in this direction could encompass developing novel model architectures, optimizing LLM training processes, and validating the efficacy and accuracy of LLM-generated test cases in practical OSF tasks.

\textbf{Dependency-Enhanced Fuzzing.} Dependency issues are a key challenge in improving the effectiveness of OSF. Current approaches use program analysis techniques to extract the control flow and data flow of the PUT to establish dependency relationships between seeds. Although~such approaches can generate semantically correct seed sequence, they often struggle to trigger deep vulnerabilities with state dependencies due to the lack of contextual information in static control flow dependencies. For example, \texttt{setsockopt} must be called twice in a row to trigger a specific~vulnerability \cite{Xu2024MOCKOK}. Additionally, while the latest approach, Mock, dynamically learns state dependencies using neural network models, it affects fuzzing efficiency and introduces a degree of randomness. To simultaneously address both control flow and state dependencies, future research could explore minimal common seed sequences by mining sources such as CVE reports (state dependencies), PoCs (state dependencies), and real-world applications (control flow dependencies) before fuzzing is executed. Moreover, by introducing a domain-specific language (DSL) in Syzlang to describe~the dependency relationships of seeds, an end-to-end dependency-enhanced OSF solution~could be achieved. Although DSL cannot cover all dependencies, it provides a solid starting point for OSF, enabling seamless integration with learning models, dynamic program analysis, and other techniques during fuzzing to continuously strengthen the expressiveness and capture of dependencies.

\textbf{Embedded Operating System Fuzzing.} With the widespread deployment of embedded OSs (\eg RT-Linux, FreeRTOS, Zephyr, \etc)~in~industrial software infrastructure, \eg autonomous driving, ensuring their security and reliability is of utmost importance. However, vulnerabilities within the code of these embedded open-source OSs may present significant long-term security and safety threats to these industries. Furthermore, the application and effectiveness of fuzzing techniques in embedded devices are significantly~constrained by the limited computational resources and the high costs of experimental equipment. While simulation environments offer a compromise between flexibility and realism, they fall short~of accurately replicating the intricate interplay of hardware and software found in embedded systems. This limitation becomes particularly acute in fuzzing OSs for intelligent vehicles, where not all aspects of embedded hardware and software can be effectively simulated. Consequently, a significant research direction for the future involves identifying and developing methods to improve both the execution efficiency and the code coverage of fuzzing techniques, specifically within the practical constraints of embedded environments.

\textbf{Rust Kernel Fuzzing.} With the increasing adoption of Rust in open-source OS kernel development due to its memory safety features, fuzzing Rust kernels has become critically important. Although Syzkaller is widely used for kernel fuzzing, its reliance on C-centric syscall descriptions (Syzlang) poses challenges when applied directly to Rust environments. Research efforts could begin~by~attempting to port Syzkaller to support Rust kernels, focusing on analyzing the relationships between syscalls. 
The Rust community is still in the process of developing features analogous to the Sanitizer and kcov modules, which are crucial for advanced fuzzing tasks. The creation of Rust equivalents for these tools, along with a syscall description language tailored for Rust, could significantly advance the field. Furthermore, by leveraging LLMs to accelerate~the~generation of automatic test cases, a higher degree of automation and efficiency can be achieved in fuzzing~Rust~kernels.



\section{Conclusions}\label{Section8}

We conduct the first systematic literature review of OSF. Overall, we classify the literature according the four OS layers, \ie kernels, file systems, drivers, and hypervisors. We first summarize the general workflow of OSF, and then elaborate the details of each step of OSF. Further, we summarize unique fuzzing challenges for different OS layers. Based on the findings from our systematic survey, we discuss the future research directions in OSF. We hope our work will encourage further research in OSF and provide valuable guidance to newcomers in this field.





\bibliographystyle{ACM-Reference-Format}
\bibliography{src/reference}

@article{Blondel2008FastUO,
  title={Fast unfolding of communities in large networks},
  author={Vincent D. Blondel and Jean-Loup Guillaume and Renaud Lambiotte and Etienne Lefebvre},
  journal={Journal of Statistical Mechanics: Theory and Experiment},
  volume={2008},
  pages={P10008},
  year={2008},
}

@inproceedings{wohlin2014guidelines,
  title={Guidelines for snowballing in systematic literature studies and a replication in software engineering},
  author={Wohlin, Claes},
  booktitle={18th international conference on evaluation and assessment in software engineering},
  pages={1--10},
  year={2014}
}

@article{garousi2016systematic,
  title={A systematic literature review of literature reviews in software testing},
  author={Garousi, Vahid and M{\"a}ntyl{\"a}, Mika V},
  journal={Information and Software Technology},
  volume={80},
  pages={195--216},
  year={2016}
}

@software{Syzkaller,
  author = {Google},
  title = {Syzkaller: an unsupervised coverage-guided kernel fuzzer},
  url = {https://github.com/google/syzkaller.},
  year = {2015}
}

@software{syzlang,
  author       = {Google},
  title        = {Syscall Descriptions Syntax},
  year         = {2024},
  url          = {https://github.com/google/syzkaller/blob/master/docs/syscall_descriptions_syntax.md}
}

@software{AFL,
  author = {lcamtuf},
  title = {American fuzzy lop},
  url = {https://lcamtuf.coredump.cx/afl/.},
  year = {2013}
}

@software{TriforceAFL,
  author = {NCC Group},
  title = {FL/QEMU Fuzzing with Full-system Emulation},
  url = {https://github.com/nccgroup/TriforceAFL},
  year = {2017}
}

@software{libfuzzer,
  author = {LLVM Project},
  title = {libFuzzer - a library for coverage-guided fuzz testing},
  url = {https://llvm.org/docs/LibFuzzer.html},
  year = {2018}
}

@software{Trinity,
  author = {D.Jones},
  title = {Linux system call fuzzer},
  url = {"https://github.com/kernelslacker/trinity,"},
  year = {2015}
}

@misc{Strace,
  title = {Strace},
  url = {https://strace.io/},
  year = {2017}
}

@misc{Wireshark,
  title = {Wireshark},
  url = {https://www.wireshark.org.},
  year = {2024}
}

@misc{usbmon,
  author       = {The kernel development community},
  title        = {usbmon},
  year         = {2021},
  howpublished = {\url{https://www.kernel.org/doc/html/v5.14/usb/usbmon.html}}
}

@software{llvm2024,
  title = {{The LLVM Compiler Infrastructure}},
  author = {{LLVM}},
  year = {2024},
  note = {Retrieved March 2024 from \url{https://llvm.org/}}
}

@misc{llvm_sancov,
  author       = {{LLVM}},
  title        = {{LLVM's SanitizerCoverage}},
  howpublished = {\url{https://clang.llvm.org/docs/SanitizerCoverage.html}},
  note         = {Accessed: Apr. 29, 2021},
}

@inproceedings{bellard2005qemu,
  title={QEMU, a fast and portable dynamic translator.},
  author={Bellard, Fabrice},
  booktitle={USENIX annual technical conference, FREENIX Track},
  pages={10--5555},
  year={2005}
}

@inproceedings{TAO2024MONARCH,
  title = {MONARCH: a fuzzing framework for distributed file systems},
  author = {Lyu, Tao and Zhang, Liyi and Feng, Zhiyao and Pan, Yueyang and Ren, Yujie and Xu, Meng and Payer, Mathias and Kashyap, Sanidhya},
  booktitle = {USENIX Conference on Usenix Annual Technical Conference},
  pages = {529--543},
  year = {2024}
}

@inproceedings{Pustogarov2020ExvivoDA,
  title={Ex-vivo dynamic analysis framework for Android device drivers},
  author={Pustogarov, Ivan and Wu, Qian and Lie, David},
  booktitle={2020 IEEE Symposium on Security and Privacy},
  pages={1088--1105},
  year={2020}
}

@inproceedings{Liu2023ViDeZZoDV,
  title={Videzzo: Dependency-aware virtual device fuzzing},
  author={Liu, Qiang and Toffalini, Flavio and Zhou, Yajin and Payer, Mathias},
  booktitle={IEEE Symposium on Security and Privacy},
  pages={3228--3245},
  year={2023}
}

@article{Hung2024BRFFT,
  title={BRF: Fuzzing the eBPF Runtime},
  author={Hung, Hsin-Wei and Amiri Sani, Ardalan},
  journal={ACM on Software Engineering},
  volume={1},
  number={FSE},
  pages={1152--1171},
  year={2024}
}

@inproceedings{Xu2024MOCKOK,
  title={Mock: optimizing kernel fuzzing mutation with context-aware dependency},
  author={Xu, Jiacheng and Zhang, Xuhong and Ji, Shouling and Tian, Yuan and Zhao, Binbin and Wang, Qinying and Cheng, Peng and Chen, Jiming},
  booktitle={Network and Distributed System Security Symposium},
  year={2024}
}

@inproceedings{Bulekov2024HYPERPILLFF,
  title={$\{$HYPERPILL$\}$: Fuzzing for Hypervisor-bugs by Leveraging the Hardware Virtualization Interface},
  author={Bulekov, Alexander and Liu, Qiang and Egele, Manuel and Payer, Mathias},
  booktitle={33rd USENIX Security Symposium},
  pages={919--935},
  year={2024}
}

@inproceedings{Lee2015F2FSAN,
  title={$\{$F2FS$\}$: A new file system for flash storage},
  author={Lee, Changman and Sim, Dongho and Hwang, Jooyoung and Cho, Sangyeun},
  booktitle={13th USENIX Conference on File and Storage Technologies},
  pages={273--286},
  year={2015}
}

@article{Rodeh2013BTRFSTL,
  title={BTRFS: The Linux B-tree filesystem},
  author={Rodeh, Ohad and Bacik, Josef and Mason, Chris},
  journal={ACM Transactions on Storage},
  volume={9},
  number={3},
  pages={1--32},
  year={2013}
}

@inproceedings{Cao2007Ext4TN,
  title={Ext4: The Next Generation of Ext2/3 Filesystem.},
  author={Cao, Mingming and Bhattacharya, Suparna and Ts'o, Ted},
  booktitle={LSF},
  year={2007}
}

@inproceedings{Hao2022DemystifyingTD,
  title={Demystifying the dependency challenge in kernel fuzzing},
  author={Hao, Yu and Zhang, Hang and Li, Guoren and Du, Xingyun and Qian, Zhiyun and Sani, Ardalan Amiri},
  booktitle={44th International Conference on Software Engineering},
  pages={659--671},
  year={2022}
}

@article{Chizpurfle2019,
  title={Evolutionary fuzzing of android OS vendor system services},
  author={Cotroneo, Domenico and Iannillo, Antonio Ken and Natella, Roberto},
  journal={Empirical Software Engineering},
  volume={24},
  pages={3630--3658},
  year={2019}
}

@inproceedings{2017DIFUZE,
  title={Difuze: Interface aware fuzzing for kernel drivers},
  author={Corina, Jake and Machiry, Aravind and Salls, Christopher and Shoshitaishvili, Yan and Hao, Shuang and Kruegel, Christopher and Vigna, Giovanni},
  booktitle={ACM SIGSAC Conference on Computer and Communications Security},
  pages={2123--2138},
  year={2017}
}

@inproceedings{chen2020koobe,
  title={$\{$KOOBE$\}$: Towards facilitating exploit generation of kernel $\{$Out-Of-Bounds$\}$ write vulnerabilities},
  author={Chen, Weiteng and Zou, Xiaochen and Li, Guoren and Qian, Zhiyun},
  booktitle={29th USENIX Security Symposium},
  pages={1093--1110},
  year={2020}
}

@inproceedings{zhao2022statefuzz,
  title={$\{$StateFuzz$\}$: System $\{$Call-Based$\}$$\{$State-Aware$\}$ linux driver fuzzing},
  author={Zhao, Bodong and Li, Zheming and Qin, Shisong and Ma, Zheyu and Yuan, Ming and Zhu, Wenyu and Tian, Zhihong and Zhang, Chao},
  booktitle={31st USENIX Security Symposium},
  pages={3273--3289},
  year={2022}
}

@inproceedings{sun2021healer,
  title={Healer: Relation learning guided kernel fuzzing},
  author={Sun, Hao and Shen, Yuheng and Wang, Cong and Liu, Jianzhong and Jiang, Yu and Chen, Ting and Cui, Aiguo},
  booktitle={ACM SIGOPS 28th Symposium on Operating Systems Principles},
  pages={344--358},
  year={2021}
}

@article{shen2021rtkaller,
  title={Rtkaller: State-aware task generation for RTOS fuzzing},
  author={Shen, Yuheng and Sun, Hao and Jiang, Yu and Shi, Heyuan and Yang, Yixiao and Chang, Wanli},
  journal={ACM Transactions on Embedded Computing Systems},
  volume={20},
  number={5s},
  pages={1--22},
  year={2021}
}

@article{shen2022tardis,
  title={Tardis: Coverage-guided embedded operating system fuzzing},
  author={Shen, Yuheng and Xu, Yiru and Sun, Hao and Liu, Jianzhong and Xu, Zichen and Cui, Aiguo and Shi, Heyuan and Jiang, Yu},
  journal={IEEE Transactions on Computer-Aided Design of Integrated Circuits and Systems},
  volume={41},
  number={11},
  pages={4563--4574},
  year={2022}
}

@inproceedings{jeong2023segfuzz,
  title={SEGFUZZ: Segmentizing Thread Interleaving to Discover Kernel Concurrency Bugs through Fuzzing},
  author={Jeong, Dae R and Lee, Byoungyoung and Shin, Insik and Kwon, Youngjin},
  booktitle={2023 IEEE Symposium on Security and Privacy},
  pages={2104--2121},
  year={2023}
}

@inproceedings{xu2020krace,
  title={Krace: Data race fuzzing for kernel file systems},
  author={Xu, Meng and Kashyap, Sanidhya and Zhao, Hanqing and Kim, Taesoo},
  booktitle={2020 IEEE Symposium on Security and Privacy},
  pages={1643--1660},
  year={2020}
}

@inproceedings{jeong2019razzer,
  title={Razzer: Finding kernel race bugs through fuzzing},
  author={Jeong, Dae R and Kim, Kyungtae and Shivakumar, Basavesh and Lee, Byoungyoung and Shin, Insik},
  booktitle={2019 IEEE Symposium on Security and Privacy},
  pages={754--768},
  year={2019}
}

@inproceedings{lin2022grebe,
  title={GREBE: Unveiling exploitation potential for Linux kernel bugs},
  author={Lin, Zhenpeng and Chen, Yueqi and Wu, Yuhang and Mu, Dongliang and Yu, Chensheng and Xing, Xinyu and Li, Kang},
  booktitle={2022 IEEE Symposium on Security and Privacy},
  pages={2078--2095},
  year={2022}
}

@inproceedings{schumilo2021nyx,
  title={Nyx: Greybox hypervisor fuzzing using fast snapshots and affine types},
  author={Schumilo, Sergej and Aschermann, Cornelius and Abbasi, Ali and W{\"o}rner, Simon and Holz, Thorsten},
  booktitle={30th USENIX Security Symposium},
  pages={2597--2614},
  year={2021}
}

@inproceedings{schumilo2020hyper,
  title={HYPER-CUBE: High-Dimensional Hypervisor Fuzzing.},
  author={Schumilo, Sergej and Aschermann, Cornelius and Abbasi, Ali and W{\"o}rner, Simon and Holz, Thorsten},
  booktitle={NDSS},
  year={2020}
}

@inproceedings{henderson2017vdf,
  title={Vdf: Targeted evolutionary fuzz testing of virtual devices},
  author={Henderson, Andrew and Yin, Heng and Jin, Guang and Han, Hao and Deng, Hongmei},
  booktitle={Research in Attacks, Intrusions, and Defenses: 20th International Symposium, RAID 2017, Atlanta, GA, USA, September 18--20, 2017, Proceedings},
  pages={3--25},
  year={2017}
}

@inproceedings{pan2021V-shuttle,
  title={V-shuttle: Scalable and semantics-aware hypervisor virtual device fuzzing},
  author={Pan, Gaoning and Lin, Xingwei and Zhang, Xuhong and Jia, Yongkang and Ji, Shouling and Wu, Chunming and Ying, Xinlei and Wang, Jiashui and Wu, Yanjun},
  booktitle={ACM SIGSAC Conference on Computer and Communications Security},
  pages={2197--2213},
  year={2021}
}

@inproceedings{myung2022mundofuzz,
  title={$\{$MundoFuzz$\}$: Hypervisor fuzzing with statistical coverage testing and grammar inference},
  author={Myung, Cheolwoo and Lee, Gwangmu and Lee, Byoungyoung},
  booktitle={31st USENIX Security Symposium},
  pages={1257--1274},
  year={2022}
}

@inproceedings{bulekov2022morphuzz,
  title={MORPHUZZ: Bending (input) space to fuzz virtual devices},
  author={Bulekov, Alexander and Das, Bandan and Hajnoczi, Stefan and Egele, Manuel},
  booktitle={31st USENIX Security Symposium},
  pages={1221--1238},
  year={2022}
}

@article{cesarano2023iris,
  title={IRIS: a Record and Replay Framework to Enable Hardware-assisted Virtualization Fuzzing},
  author={Cesarano, Carmine and Cinque, Marcello and Cotroneo, Domenico and De Simone, Luigi and Farina, Giorgio},
  journal={arXiv preprint arXiv:2303.12817},
  year={2023}
}

@inproceedings{CONZZER2022context,
  title={Context-sensitive and directional concurrency fuzzing for data-race detection},
  author={Jiang, Zu-Ming and Bai, Jia-Ju and Lu, Kangjie and Hu, Shi-Min},
  booktitle={Network and Distributed Systems Security Symposium 2022},
  year={2022}
}

@article{Hydra2020finding,
  title={Finding bugs in file systems with an extensible fuzzing framework},
  author={Kim, Seulbae and Xu, Meng and Kashyap, Sanidhya and Yoon, Jungyeon and Xu, Wen and Kim, Taesoo},
  journal={ACM Transactions on Storage},
  volume={16},
  number={2},
  pages={1--35},
  year={2020}
}

@inproceedings{JANUS2019fuzzing,
  title={Fuzzing file systems via two-dimensional input space exploration},
  author={Xu, Wen and Moon, Hyungon and Kashyap, Sanidhya and Tseng, Po-Ning and Kim, Taesoo},
  booktitle={2019 IEEE Symposium on Security and Privacy},
  pages={818--834},
  year={2019}
}

@inproceedings{bulekov2023FUZZNG,
  title={No grammar, no problem: Towards fuzzing the linux kernel without system-call descriptions},
  author={Bulekov, Alexander and Das, Bandan and Hajnoczi, Stefan and Egele, Manuel},
  booktitle={Network and Distributed System Security Symposium},
  year={2023}
}

@inproceedings{fleischer2023actor,
  title={$\{$ACTOR$\}$:$\{$Action-Guided$\}$ Kernel Fuzzing},
  author={Fleischer, Marius and Das, Dipanjan and Bose, Priyanka and Bai, Weiheng and Lu, Kangjie and Payer, Mathias and Kruegel, Christopher and Vigna, Giovanni},
  booktitle={32nd USENIX Security Symposium},
  pages={5003--5020},
  year={2023}
}

@inproceedings{chen2022sfuzz,
  title={SFuzz: Slice-based Fuzzing for Real-Time Operating Systems},
  author={Chen, Libo and Cai, Quanpu and Ma, Zhenbang and Wang, Yanhao and Hu, Hong and Shen, Minghang and Liu, Yue and Guo, Shanqing and Duan, Haixin and Jiang, Kaida and others},
  booktitle={ACM SIGSAC Conference on Computer and Communications Security},
  pages={485--498},
  year={2022}
}

@inproceedings{peng2020usbfuzz,
  title={$\{$USBFuzz$\}$: A Framework for Fuzzing $\{$USB$\}$ Drivers by Device Emulation},
  author={Peng, Hui and Payer, Mathias},
  booktitle={29th USENIX Security Symposium},
  pages={2559--2575},
  year={2020}
}

@inproceedings{hao2023syzdescribe,
  title={Syzdescribe: Principled, automated, static generation of syscall descriptions for kernel drivers},
  author={Hao, Yu and Li, Guoren and Zou, Xiaochen and Chen, Weiteng and Zhu, Shitong and Qian, Zhiyun and Sani, Ardalan Amiri},
  booktitle={2023 IEEE Symposium on Security and Privacy},
  pages={3262--3278},
  year={2023}
}

@inproceedings{wu2023devfuzz,
  title={DEVFUZZ: automatic device model-guided device driver fuzzing},
  author={Wu, Yilun and Zhang, Tong and Jung, Changhee and Lee, Dongyoon},
  booktitle={2023 IEEE Symposium on Security and Privacy},
  pages={3246--3261},
  year={2023}
}

@inproceedings{zhao2022semantic,
  title={Semantic-informed driver fuzzing without both the hardware devices and the emulators},
  author={Zhao, Wenjia and Lu, Kangjie and Wu, Qiushi and Qi, Yong},
  booktitle={Network and Distributed Systems Security Symposium 2022},
  year={2022}
}

@inproceedings{ma2022printfuzz,
  title={Printfuzz: Fuzzing linux drivers via automated virtual device simulation},
  author={Ma, Zheyu and Zhao, Bodong and Ren, Letu and Li, Zheming and Ma, Siqi and Luo, Xiapu and Zhang, Chao},
  booktitle={31st ACM SIGSOFT International Symposium on Software Testing and Analysis},
  pages={404--416},
  year={2022}
}

@inproceedings{maier2021bsod,
  title={BSOD: Binary-only scalable fuzzing of device drivers},
  author={Maier, Dominik and Toepfer, Fabian},
  booktitle={24th International Symposium on Research in Attacks, Intrusions and Defenses},
  pages={48--61},
  year={2021}
}

@inproceedings{song2020agamotto,
  title={Agamotto: Accelerating kernel driver fuzzing with lightweight virtual machine checkpoints},
  author={Song, Dokyung and Hetzelt, Felicitas and Kim, Jonghwan and Kang, Brent Byunghoon and Seifert, Jean-Pierre and Franz, Michael},
  booktitle={29th USENIX Security Symposium},
  pages={2541--2557},
  year={2020}
}

@inproceedings{shen2022drifuzz,
  title={Drifuzz: Harvesting bugs in device drivers from golden seeds},
  author={Shen, Zekun and Roongta, Ritik and Dolan-Gavitt, Brendan},
  booktitle={31st USENIX Security Symposium},
  pages={1275--1290},
  year={2022}
}

@inproceedings{sun2022ksg,
  title={$\{$KSG$\}$: Augmenting kernel fuzzing with system call specification generation},
  author={Sun, Hao and Shen, Yuheng and Liu, Jianzhong and Xu, Yiru and Jiang, Yu},
  booktitle={2022 USENIX Annual Technical Conference},
  pages={351--366},
  year={2022}
}

@inproceedings{zou2022syzscope,
  title={$\{$Syzscope$\}$: Revealing $\{$high-risk$\}$ security impacts of $\{$fuzzer-exposed$\}$ bugs in linux kernel},
  author={Zou, Xiaochen and Li, Guoren and Chen, Weiteng and Zhang, Hang and Qian, Zhiyun},
  booktitle={31st USENIX Security Symposium},
  pages={3201--3217},
  year={2022}
}

@inproceedings{wang2021syzvegas,
  title={$\{$SyzVegas$\}$: Beating kernel fuzzing odds with reinforcement learning},
  author={Wang, Daimeng and Zhang, Zheng and Zhang, Hang and Qian, Zhiyun and Krishnamurthy, Srikanth V and Abu-Ghazaleh, Nael},
  booktitle={30th USENIX Security Symposium},
  pages={2741--2758},
  year={2021}
}

@inproceedings{kim2020hfl,
  title={HFL: Hybrid Fuzzing on the Linux Kernel.},
  author={Kim, Kyungtae and Jeong, Dae R and Kim, Chung Hwan and Jang, Yeongjin and Shin, Insik and Lee, Byoungyoung},
  booktitle={NDSS},
  year={2020}
}

@inproceedings{liang2020xafl,
  title={X-afl: A kernel fuzzer combining passive and active fuzzing},
  author={Liang, Hongliang and Chen, Yixiu and Xie, Zhuosi and Liang, Zhiyi},
  booktitle={13th European workshop on Systems Security},
  pages={13--18},
  year={2020}
}

@inproceedings{chen2019slake,
  title={Slake: Facilitating slab manipulation for exploiting vulnerabilities in the linux kernel},
  author={Chen, Yueqi and Xing, Xinyu},
  booktitle={ACM SIGSAC Conference on Computer and Communications Security},
  pages={1707--1722},
  year={2019}
}

@inproceedings{shi2019industry,
  title={Industry practice of coverage-guided enterprise linux kernel fuzzing},
  author={Shi, Heyuan and Wang, Runzhe and Fu, Ying and Wang, Mingzhe and Shi, Xiaohai and Jiao, Xun and Song, Houbing and Jiang, Yu and Sun, Jiaguang},
  booktitle={27th ACM Joint Meeting on European Software Engineering Conference and Symposium on the Foundations of Software Engineering},
  pages={986--995},
  year={2019}
}

@inproceedings{tan2023syzdirect,
  title={Syzdirect: Directed greybox fuzzing for linux kernel},
  author={Tan, Xin and Zhang, Yuan and Lu, Jiadong and Xiong, Xin and Liu, Zhuang and Yang, Min},
  booktitle={ACM SIGSAC Conference on Computer and Communications Security},
  pages={1630--1644},
  year={2023}
}

@inproceedings{schumilo2017kafl,
  title={$\{$kAFL$\}$:$\{$Hardware-Assisted$\}$ feedback fuzzing for $\{$OS$\}$ kernels},
  author={Schumilo, Sergej and Aschermann, Cornelius and Gawlik, Robert and Schinzel, Sebastian and Holz, Thorsten},
  booktitle={26th USENIX security symposium},
  pages={167--182},
  year={2017}
}

@inproceedings{you2017semfuzz,
  title={Semfuzz: Semantics-based automatic generation of proof-of-concept exploits},
  author={You, Wei and Zong, Peiyuan and Chen, Kai and Wang, XiaoFeng and Liao, Xiaojing and Bian, Pan and Liang, Bin},
  booktitle={ACM SIGSAC conference on computer and communications security},
  pages={2139--2154},
  year={2017}
}

@inproceedings{pailoor2018moonshine,
  title={$\{$MoonShine$\}$: Optimizing $\{$OS$\}$ fuzzer seed selection with trace distillation},
  author={Pailoor, Shankara and Aday, Andrew and Jana, Suman},
  booktitle={27th USENIX Security Symposium},
  pages={729--743},
  year={2018}
}

@inproceedings{wu2018fuze,
  title={$\{$FUZE$\}$: Towards facilitating exploit generation for kernel $\{$Use-After-Free$\}$ vulnerabilities},
  author={Wu, Wei and Chen, Yueqi and Xu, Jun and Xing, Xinyu and Gong, Xiaorui and Zou, Wei},
  booktitle={27th USENIX Security Symposium},
  pages={781--797},
  year={2018}
}

@inproceedings{schwarz2018automated,
  title={Automated detection, exploitation, and elimination of double-fetch bugs using modern cpu features},
  author={Schwarz, Michael and Gruss, Daniel and Lipp, Moritz and Maurice, Cl{\'e}mentine and Schuster, Thomas and Fogh, Anders and Mangard, Stefan},
  booktitle={Asia Conference on Computer and Communications Security},
  pages={587--600},
  year={2018}
}

@article{yang2023kernelgpt,
  title={KernelGPT: Enhanced Kernel Fuzzing via Large Language Models},
  author={Yang, Chenyuan and Zhao, Zijie and Zhang, Lingming},
  journal={arXiv preprint arXiv:2401.00563},
  year={2023}
}

@inproceedings{Unicorefuzz2019,
  title={Unicorefuzz: On the viability of emulation for kernelspace fuzzing},
  author={Maier, Dominik and Radtke, Benedikt and Harren, Bastian},
  booktitle={13th USENIX Workshop on Offensive Technologies},
  year={2019}
}

@inproceedings{Jang2023ReUSB,
  title={ReUSB: Replay-Guided USB Driver Fuzzing.},
  author={Jang, Jisoo and Kang, Minsuk and Song, Dokyung},
  booktitle={USENIX Security Symposium},
  pages={2921--2938},
  year={2023}
}

@inproceedings{Yuan2023DDRace,
  title={DDRace: Finding Concurrency UAF Vulnerabilities in Linux Drivers with Directed Fuzzing.},
  author={Yuan, Ming and Zhao, Bodong and Li, Penghui and Liang, Jiashuo and Han, Xinhui and Luo, Xiapu and Zhang, Chao},
  booktitle={USENIX Security Symposium},
  pages={2849--2866},
  year={2023}
}

@inproceedings{Liu2023LFuzz,
  title={LFuzz: Exploiting Locality-Enabled Techniques for File-System Fuzzing},
  author={Liu, Wenqing and Wang, An-I Andy},
  booktitle={European Symposium on Research in Computer Security},
  pages={507--525},
  year={2023}
}

@inproceedings{Liu2023VDGuard,
  title={VD-Guard: DMA Guided Fuzzing for Hypervisor Virtual Device},
  author={Liu, Yuwei and Chen, Siqi and Xie, Yuchong and Wang, Yanhao and Chen, Libo and Wang, Bin and Zeng, Yingming and Xue, Zhi and Su, Purui},
  booktitle={2023 38th IEEE/ACM International Conference on Automated Software Engineering},
  pages={1676--1687},
  year={2023}
}

@article{Yun2022FuzzingOE,
  title={Fuzzing of embedded systems: A survey},
  author={Yun, Joobeom and Rustamov, Fayozbek and Kim, Juhwan and Shin, Youngjoo},
  journal={ACM Computing Surveys},
  volume={55},
  number={7},
  pages={1--33},
  year={2022}
}

@article{Zhu2022FuzzingAS,
  title={Fuzzing: a survey for roadmap},
  author={Zhu, Xiaogang and Wen, Sheng and Camtepe, Seyit and Xiang, Yang},
  journal={ACM Computing Surveys},
  volume={54},
  number={11s},
  pages={1--36},
  year={2022}
}

@article{Mans2018TheAS,
  title={The art, science, and engineering of fuzzing: A survey},
  author={Man{\`e}s, Valentin JM and Han, HyungSeok and Han, Choongwoo and Cha, Sang Kil and Egele, Manuel and Schwartz, Edward J and Woo, Maverick},
  journal={IEEE Transactions on Software Engineering},
  volume={47},
  number={11},
  pages={2312--2331},
  year={2019}
}

@article{Li2018FuzzingAS,
  title={Fuzzing: a survey},
  author={Li, Jun and Zhao, Bodong and Zhang, Chao},
  journal={Cybersecurity},
  volume={1},
  pages={1--13},
  year={2018}
}

@article{Liang2018FuzzingSO,
  title={Fuzzing: State of the art},
  author={Liang, Hongliang and Pei, Xiaoxiao and Jia, Xiaodong and Shen, Wuwei and Zhang, Jian},
  journal={IEEE Transactions on Reliability},
  volume={67},
  number={3},
  pages={1199--1218},
  year={2018}
}

@article{Saavedra2019ARO,
  title={A Review of Machine Learning Applications in Fuzzing},
  author={Gary J. Saavedra and Kathryn N. Rodhouse and Daniel M. Dunlavy and Philip W Kegelmeyer},
  journal={ArXiv},
  year={2019},
  volume={abs/1906.11133}
}

@article{Wang2020SoKTP,
  title={SoK: The Progress, Challenges, and Perspectives of Directed Greybox Fuzzing},
  author={Pengfei Wang and Xu Zhou},
  journal={ArXiv},
  year={2020},
  volume={abs/2005.11907}
}

@article{Godefroid2020,
  title={Fuzzing: Hack, art, and science},
  author={Godefroid, Patrice},
  journal={Communications of the ACM},
  volume={63},
  number={2},
  pages={70--76},
  year={2020}
}

@article{bohme2020fuzzing,
  title={Fuzzing: Challenges and reflections},
  author={B{\"o}hme, Marcel and Cadar, Cristian and Roychoudhury, Abhik},
  journal={IEEE Software},
  volume={38},
  number={3},
  pages={79--86},
  year={2020}
}

@article{Wang2019ASR,
  title={A systematic review of fuzzing based on machine learning techniques},
  author={Wang, Yan and Jia, Peng and Liu, Luping and Huang, Cheng and Liu, Zhonglin},
  journal={PloS one},
  volume={15},
  number={8},
  pages={e0237749},
  year={2020}
}

@inproceedings{Zhang2018SurveyOD,
  title={Survey of directed fuzzy technology},
  author={Zhang, Yan and Zhang, Junwen and Zhang, Dalin and Mu, Yongmin},
  booktitle={2018 IEEE 9th International Conference on Software Engineering and Service Science},
  pages={1--4},
  year={2018}
}

@article{Eisele2022EmbeddedFA,
  title={Embedded fuzzing: a review of challenges, tools, and solutions},
  author={Eisele, Max and Maugeri, Marcello and Shriwas, Rachna and Huth, Christopher and Bella, Giampaolo},
  journal={Cybersecurity},
  volume={5},
  number={1},
  pages={18},
  year={2022}
}

@article{Mallissery2023DemystifyTF,
  title={Demystify the fuzzing methods: A comprehensive survey},
  author={Mallissery, Sanoop and Wu, Yu-Sung},
  journal={ACM Computing Surveys},
  volume={56},
  number={3},
  pages={1--38},
  year={2023}
}

@inproceedings{Xu2024Saturn,
  title={Saturn: Host-Gadget Synergistic USB Driver Fuzzing}, 
  author={Xu, Yiru and Sun, Hao and Liu, Jianzhong and Shen, Yuheng and Jiang, Yu},
  booktitle={2024 IEEE Symposium on Security and Privacy}, 
  year={2024}
}

@inproceedings{Huster2024ToBoldly,
  title={To Boldly Go Where No Fuzzer Has Gone Before: Finding Bugs in Linux’ Wireless Stacks through VirtIO Devices}, 
  author={Huster, Sönke and Hollick, Matthias and Classen, Jiska},
  booktitle={2024 IEEE Symposium on Security and Privacy}, 
  year={2024}
}

@inproceedings{Chen2024SyzGen++,
  title = {SyzGen++: Dependency Inference for Augmenting Kernel Driver Fuzzing},
  author = {Chen, Weiteng and Hao, Yu and Zhang, Zheng and Zou, Xiaochen and Kirat, Dhilung and Mishra, Shachee and Schales, Douglas and Jang, Jiyong and Qian, Zhiyun},
  booktitle = {2024 IEEE Symposium on Security and Privacy},
  year = {2024}
}

@misc{NVD,
  title        = {National Vulnerability Database},
  howpublished = {\url{https://nvd.nist.gov/vuln/search}},
  note         = {Accessed: 2024-09-14},
  year         = {2024},
  author       = {{National Institute of Standards and Technology (NIST)}}
}

@article{2006Intel6,
  title={Intel{\textregistered} 64 and ia-32 architectures software developer’s manual},
  author={Guide, Part},
  journal={Volume 3B: System programming Guide, Part},
  volume={2},
  number={11},
  pages={0--40},
  year={2011}
}

@article{Yang2023WhiteboxCF,
  title={White-box Compiler Fuzzing Empowered by Large Language Models},
  author={Chenyuan Yang and Yinlin Deng and Runyu Lu and Jiayi Yao and Jiawei Liu and Reyhaneh Jabbarvand and Lingming Zhang},
  journal={ArXiv},
  year={2023},
  volume={abs/2310.15991}
}

@inproceedings{Holler2012FuzzingWC,
  title={Fuzzing with code fragments},
  author={Holler, Christian and Herzig, Kim and Zeller, Andreas},
  booktitle={21st USENIX Security Symposium},
  pages={445--458},
  year={2012}
}

@article{Miller1990AnES,
  title={An empirical study of the reliability of UNIX utilities},
  author={Miller, Barton P and Fredriksen, Lars and So, Bryan},
  journal={Communications of the ACM},
  volume={33},
  number={12},
  pages={32--44},
  year={1990}
}

@inproceedings{Zhang2015AndroidRA,
  title={Android root and its providers: A double-edged sword},
  author={Zhang, Hang and She, Dongdong and Qian, Zhiyun},
  booktitle={22nd ACM SIGSAC Conference on Computer and Communications Security},
  pages={1093--1104},
  year={2015}
}

@inproceedings{Xu2015FromCT,
  title={From collision to exploitation: Unleashing use-after-free vulnerabilities in linux kernel},
  author={Xu, Wen and Li, Juanru and Shu, Junliang and Yang, Wenbo and Xie, Tianyi and Zhang, Yuanyuan and Gu, Dawu},
  booktitle={ACM SIGSAC Conference on Computer and Communications Security},
  pages={414--425},
  year={2015}
}

@inproceedings{Song2019PeriScopeAE,
  title={Periscope: An effective probing and fuzzing framework for the hardware-os boundary},
  author={Song, Dokyung and Hetzelt, Felicitas and Das, Dipanjan and Spensky, Chad and Na, Yeoul and Volckaert, Stijn and Vigna, Giovanni and Kruegel, Christopher and Seifert, Jean-Pierre and Franz, Michael},
  booktitle={NDSS},
  year={2019}
}

@inproceedings{Lu2013ASO,
  title={A study of linux file system evolution},
  author={Lu, Lanyue and Arpaci-Dusseau, Andrea C and Arpaci-Dusseau, Remzi H and Lu, Shan},
  booktitle={11th USENIX Conference on file and storage technologies},
  pages={31--44},
  year={2013}
}

@article{UBsan,
  title={Understanding integer overflow in C/C++},
  author={Dietz, Will and Li, Peng and Regehr, John and Adve, Vikram},
  journal={ACM Transactions on Software Engineering and Methodology},
  volume={25},
  number={1},
  pages={1--29},
  year={2015}
}

@inproceedings{happensbefore,
  title={Diagnosing kernel concurrency failures with AITIA},
  author={Jeong, Dae R and Jung, Minkyu and Lee, Yoochan and Lee, Byoungyoung and Shin, Insik and Kwon, Youngjin},
  booktitle={Eighteenth European Conference on Computer Systems},
  pages={94--110},
  year={2023}
}

@inproceedings{Vinesh2019ConFuzzACF,
  title={Confuzz—a concurrency fuzzer},
  author={Vinesh, Nischai and Sethumadhavan, M},
  booktitle={First International Conference on Sustainable Technologies for Computational Intelligence},
  pages={667--691},
  year={2020}
}

@inproceedings{Johansson2018RandomTW,
  title={Random testing with sanitizers to detect concurrency bugs in embedded avionics software},
  author={Viktor Johansson and A.L.M. Vallen},
  year={2018}
}

@inproceedings{Chen2020MUZZTG,
  title={$\{$MUZZ$\}$: Thread-aware grey-box fuzzing for effective bug hunting in multithreaded programs},
  author={Chen, Hongxu and Guo, Shengjian and Xue, Yinxing and Sui, Yulei and Zhang, Cen and Li, Yuekang and Wang, Haijun and Liu, Yang},
  booktitle={29th USENIX Security Symposium},
  pages={2325--2342},
  year={2020}
}

@inproceedings{ThreadSanitizerDR,
  title={ThreadSanitizer: data race detection in practice},
  author={Serebryany, Konstantin and Iskhodzhanov, Timur},
  booktitle={the workshop on binary instrumentation and applications},
  pages={62--71},
  year={2009}
}

@article{Savage1997EraserAD,
  title={Eraser: A dynamic data race detector for multithreaded programs},
  author={Savage, Stefan and Burrows, Michael and Nelson, Greg and Sobalvarro, Patrick and Anderson, Thomas},
  journal={ACM Transactions on Computer Systems},
  volume={15},
  number={4},
  pages={391--411},
  year={1997}
}

@inproceedings{Chou2001AnES,
  title={An empirical study of operating systems errors},
  author={Chou, Andy and Yang, Junfeng and Chelf, Benjamin and Hallem, Seth and Engler, Dawson},
  booktitle={eighteenth ACM symposium on Operating systems principles},
  pages={73--88},
  year={2001}
}

@inproceedings{Palix2011FaultsIL,
  title={Faults in Linux: Ten years later},
  author={Palix, Nicolas and Thomas, Ga{\"e}l and Saha, Suman and Calv{\`e}s, Christophe and Lawall, Julia and Muller, Gilles},
  booktitle={sixteenth international conference on Architectural support for programming languages and operating systems},
  pages={305--318},
  year={2011}
}

@inproceedings{Stoep2018AndroidSecurity,
  author = {J. V. Stoep and S. Tolvanen},
  title = {Year in review: Android kernel security},
  booktitle = {Linux Security Summit},
  year = {2018}
}

@misc{Beniamini2017Part1,
  author = {Gal Beniamini},
  title = {Over the air: Exploiting Broadcom’s Wi-Fi stack (part 1)},
  year = {2017},
  note = {\url{https://googleprojectzero.blogspot.com/2017/04/over-air-exploiting-broadcoms-wi-fi_4.html}}
}

@misc{Beniamini2017Part2,
  author = {Gal Beniamini},
  title = {Over the air: Exploiting Broadcom’s Wi-Fi stack (part 2)},
  year = {2017},
  note = {\url{https://googleprojectzero.blogspot.com/2017/04/over-air-exploiting-broadcoms-wi-fi_11.html}}
}

@misc{Chang2017,
  author = {Omri Chang},
  title = {Attacking the Windows NVIDIA driver},
  year = {2017},
  note = {\url{https://googleprojectzero.blogspot.com/2017/02/attacking-windows-nvidia-driver.html}}
}

@article{Davis2011,
  title={USB-undermining security barriers},
  author={Davis, Andy},
  journal={Black Hat Briefings},
  year={2011}
}

@article{NohlLell2014,
  title={BadUSB-On Accessories That Turn Evil},
  author={Nohl, Karsten},
  journal={Black Hat USA},
  year={2014}
}

@article{Cinque2021VirtualizingMS,
  title={Virtualizing mixed-criticality systems: A survey on industrial trends and issues},
  author={Cinque, Marcello and Cotroneo, Domenico and De Simone, Luigi and Rosiello, Stefano},
  journal={Future Generation Computer Systems},
  volume={129},
  pages={315--330},
  year={2022}
}

@manual{RTCA_DO_178C,
  title = {{DO-178C Software Considerations in Airborne Systems and Equipment Certification}},
  organization = {RTCA},
  year = {1992},
  note = {Requirements and Technical Concepts for Aviation},
}

@manual{ISO_26262_2011,
  title = {Road Vehicles - Functional Safety},
  organization = {International Organization for Standardization},
  year = {2011},
  number = {ISO 26262},
  address = {Geneva, Switzerland},
}

@article{Popek1974FormalRF,
  title={Formal requirements for virtualizable third generation architectures},
  author={Popek, Gerald J and Goldberg, Robert P},
  journal={Communications of the ACM},
  volume={17},
  number={7},
  pages={412--421},
  year={1974}
}

@article{Cilardo2021VirtualizationOM,
  title={Virtualization over multiprocessor systems-on-chip: An enabling paradigm for the industrial internet of things},
  author={Cilardo, Alessandro and Cinque, Marcello and De Simone, Luigi and Mazzocca, Nicola},
  journal={Computer},
  volume={55},
  number={10},
  pages={35--47},
  year={2022}
}

@article{Instrumentation_Huang1978,
  title={Program instrumentation and software testing},
  author={Huang, JC},
  journal={Computer},
  volume={11},
  number={4},
  pages={25--32},
  year={1978}
}

@misc{cve,
    title = {Common Vulnerabilities and Exposures},
    year = {2024},
    url = {https://cve.mitre.org},
    note = {(2024)}
}

@misc{linuxkernel,
    title = {Linux Kernel Git Repositories},
    year = {2024},
    url = {https://git.kernel.org},
    note = {(2024)}
}

@article{Brin1998PageRank,
  title={The anatomy of a large-scale hypertextual web search engine},
  author={Brin, Sergey and Page, Lawrence},
  journal={Computer networks and ISDN systems},
  volume={30},
  number={1-7},
  pages={107--117},
  year={1998}
}

@inproceedings{Bx00F6hme2016CoverageBasedGF,
  title={Coverage-based greybox fuzzing as markov chain},
  author={B{\"o}hme, Marcel and Pham, Van-Thuan and Roychoudhury, Abhik},
  booktitle={ACM SIGSAC Conference on Computer and Communications Security},
  pages={1032--1043},
  year={2016}
}

@inproceedings{Lemieux2017FairFuzzAT,
  title={Fairfuzz: A targeted mutation strategy for increasing greybox fuzz testing coverage},
  author={Lemieux, Caroline and Sen, Koushik},
  booktitle={33rd ACM/IEEE international conference on automated software engineering},
  pages={475--485},
  year={2018}
}

@inproceedings{Huang2022BEACONDG,
  title={Beacon: Directed grey-box fuzzing with provable path pruning},
  author={Huang, Heqing and Guo, Yiyuan and Shi, Qingkai and Yao, Peisen and Wu, Rongxin and Zhang, Charles},
  booktitle={2022 IEEE Symposium on Security and Privacy},
  pages={36--50},
  year={2022}
}

@inproceedings{Zong2020FuzzGuardFO,
  title={$\{$FuzzGuard$\}$: Filtering out unreachable inputs in directed grey-box fuzzing through deep learning},
  author={Zong, Peiyuan and Lv, Tao and Wang, Dawei and Deng, Zizhuang and Liang, Ruigang and Chen, Kai},
  booktitle={29th USENIX security symposium},
  pages={2255--2269},
  year={2020}
}

@inproceedings{Sutton2007FuzzingBF,
  title={Fuzzing: Brute Force Vulnerability Discovery},
  author={Michael S. Sutton and Adam R. Greene and Pedram Fardad Amini},
  year={2007}
}

@inproceedings{Koopman1997ComparingOS,
  title={Comparing operating systems using robustness benchmarks},
  author={Koopman, Philip and Sung, John and Dingman, Christopher and Siewiorek, Daniel and Marz, Ted},
  booktitle={16th IEEE Symposium on Reliable Distributed Systems},
  pages={72--79},
  year={1997}
}

@inproceedings{Muench2018Avatar2AM,
  title={Avatar 2: A multi-target orchestration platform},
  author={Muench, Marius and Nisi, Dario and Francillon, Aur{\'e}lien and Balzarotti, Davide},
  booktitle={Proc. Workshop Binary Anal. Res.},
  volume={18},
  pages={1--11},
  year={2018}
}

@inproceedings{Song2018SoKSF,
  title={SoK: Sanitizing for security},
  author={Song, Dokyung and Lettner, Julian and Rajasekaran, Prabhu and Na, Yeoul and Volckaert, Stijn and Larsen, Per and Franz, Michael},
  booktitle={2019 IEEE Symposium on Security and Privacy},
  pages={1275--1295},
  year={2019}
}

@inproceedings{2018B3,
title = {Finding crash-consistency bugs with bounded black-box crash testing},
author = {Mohan, Jayashree and Martinez, Ashlie and Ponnapalli, Soujanya and Raju, Pandian and Chidambaram, Vijay},
booktitle = {13th USENIX Conference on Operating Systems Design and Implementation},
pages = {33–50},
year = {2018}
}

@inproceedings{SibylFS2015,
title = {SibylFS: formal specification and oracle-based testing for POSIX and real-world file systems},
author = {Ridge, Tom and Sheets, David and Tuerk, Thomas and Giugliano, Andrea and Madhavapeddy, Anil and Sewell, Peter},
booktitle = {25th Symposium on Operating Systems Principles},
pages = {38–53},
year = {2015}
}

@article{Serebryany2017OSSFuzzG,
  title={$\{$OSS-Fuzz$\}$-Google's continuous fuzzing service for open source software},
  author={Serebryany, Kostya},
  year={2017}
}

@misc{Syzbot,
  author = {Google},
  title = {Syzbot},
  year = {2024},
  note = {\url{https://syzkaller.appspot.com/upstream}}
}

@misc{fstests,
  author = {Silicon Graphics Inc. (SGI)},
  title = {(x)fstests is a filesystem testing suite},
  year = {2018},
  note = {\url{https://github.com/kdave/xfstests}}
}

@misc{XFS2018,
  author = {Silicon Graphics Inc. (SGI) and Red Hat Inc},
  title = {XFS},
  year = {2018},
  note = {\url{http://xfs.org}}
}

@misc{lockdep,
  author = {Linux},
  title = {Runtime locking correctness validator},
  year = {2022},
  note = {\url{https://www.kernel.org/doc/Documentation/locking/lockdep-design.txt}}
}

@misc{KCSAN,
  title = {KCSAN: concurrency sanitizer for the Linux kernel},
  howpublished = {\url{https://google.github.io/kernel-sanitizers/KCSAN.html}},
  year = {2020}
}

@misc{TSan,
  title = {ThreadSanitizer: a data race detector for C/C++},
  howpublished = {\url{https://github.com/google/sanitizers/wiki/ThreadSanitizerCppManual}},
  year = {2020}
}

@misc{KernelAddressSanitizer2019,
  title = {Kernel AddressSanitizer},
  howpublished = {\url{https://www.kernel.org/doc/html/v4.14/dev-tools/kasan.html}},
  year = {2019}
}

@misc{MemorySanitizer2020,
  title = {MemorySanitizer},
  howpublished = {\url{https://github.com/google/sanitizers/wiki/MemorySanitizer}},
  year = {2020}
}

@misc{coresight2017,
  author = {{ARM}},
  title = {CoreSight on-chip trace and debug},
  month = may,
  year = {2017},
  note = {[Online]. Available: \url{http://infocenter.arm.com/help/index.jsp?topic=/com.arm.doc.set.coresight/index.html}},
}

@misc{simplept,
  author = {A. Kleen},
  title = {simple-pt: Simple Intel CPU processor tracing on Linux},
  howpublished = {\url{https://github.com/andikleen/simple-pt}},
}

@inproceedings{kleen2015,
  author = {A. Kleen and B. Strong},
  title = {Intel Processor Trace on Linux},
  booktitle = {Tracing Summit 2015},
  year = {2015},
}

@misc{ibm2020,
  author = {{IBM}},
  title = {Source Code Instrumentation Overview},
  howpublished = {\url{https://www.ibm.com/support/knowledgecenter/SSSHUF_8.0.0/com.ibm.rational.testrt.doc/topics/cinstruovw.html}},
  note = {Retrieved March 7, 2020},
  year = {n.d.}
}

@inproceedings{Yarom2014FLUSHRELOADAH,
  title={$\{$FLUSH+ RELOAD$\}$: A high resolution, low noise, l3 cache $\{$Side-Channel$\}$ attack},
  author={Yarom, Yuval and Falkner, Katrina},
  booktitle={23rd USENIX security symposium},
  pages={719--732},
  year={2014}
}

@inproceedings{2011IS,
  title={Cache games--bringing access-based cache attacks on AES to practice},
  author={Gullasch, David and Bangerter, Endre and Krenn, Stephan},
  booktitle={2011 IEEE Symposium on Security and Privacy},
  pages={490--505},
  year={2011}
}

@inproceedings{ASAN,
  title={$\{$AddressSanitizer$\}$: A fast address sanity checker},
  author={Serebryany, Konstantin and Bruening, Derek and Potapenko, Alexander and Vyukov, Dmitriy},
  booktitle={2012 USENIX annual technical conference},
  pages={309--318},
  year={2012}
}

@software{gcov,
  title = {{gcov: GCC Coverage}},
  author = {{GNU Project}},
  note = {Retrieved May 2024 from \url{https://gcc.gnu.org/onlinedocs/gcc/Gcov.html}}
}

@software{vyukov2018kcov,
  author = {Dmitry Vyukov},
  title = {kernel: add kcov code coverage},
  year = {2018},
  url = {https://lwn.net/Articles/671640/},
}

@software{kernel2018kasan,
  author = {{Linux kernel document}},
  title = {KernelAddressSanitizer},
  year = {2018},
  url = {https://github.com/google/kasan/wiki}
}

@software{QTest,
  title = {Features/QTest},
  url = {http://wiki.qemu.org/Features/QTest},
  year = {2024}
}

\end{document}